\documentclass[journal]{IEEEtran}
\usepackage{inputenc}
\usepackage{fontenc}
\usepackage{mathtext}
\usepackage{textcase}
\usepackage{bm}
\usepackage{amsfonts,amssymb}
\usepackage{subfigure}
\usepackage{mathrsfs}
\usepackage{color}
\usepackage[ruled]{algorithm2e}
\usepackage{textcomp}
\usepackage{multirow}
\usepackage{booktabs}

\usepackage{cite}

\ifCLASSINFOpdf
  \usepackage[pdftex]{graphicx}
\else
   \usepackage[dvips]{graphicx}
   
\fi
\usepackage{epstopdf}
\usepackage{caption}
\usepackage{amsmath}
\usepackage{algorithmic}

\usepackage{array}
\begin{document}
%
\title{Rate-Splitting Multiple Access for Multi-Antenna Joint Radar and Communications}
%
%
%

\author{Chengcheng Xu,
Bruno Clerckx, \IEEEmembership{Senior Member, IEEE},
Shiwa Chen, 
Yijie Mao, \IEEEmembership{Member, IEEE}, and
Jianyun Zhang, \IEEEmembership{Member, IEEE}
\thanks{This paper was presented in part at 2020 IEEE International Conference on Communications Workshops (ICC Workshops), Dublin, Ireland, June 2020 \cite{Xu2020ICC}.}
\thanks{Chengcheng Xu, Shiwa Chen and Jianyun Zhang are with the College of Electronic Engineering, National University of Defense Technology, Hefei 230037, China (email: xuchengcheng17@nudt.edu.cn; chenshiwa17@nudt.edu.cn; zjy.eei@163.com).}
\thanks{Bruno Clerckx and Yijie Mao are with the Communications and Signal Processing Group, Department of Electrical and Electronic Engineering, Imperial College London, London SW7 2AZ, U.K. (email: b.clerckx@imperial.ac.uk; y.mao16@imperial.ac.uk)}}

\maketitle

\begin{abstract}
Dual-Functional Radar-Communication (DFRC) system is an essential and promising technique to achieve Integrated Sensing and Communication (ISAC) for beyond 5G. In this work, we propose a powerful and unified multi-antenna DFRC transmission framework, where an additional radar sequence is transmitted apart from communication streams to enhance radar beampattern matching capability, and Rate-Splitting Multiple Access (RSMA) is adopted to better manage the interference. RSMA relies on multi-antenna Rate-Splitting (RS) with Successive Interference Cancellation (SIC) receivers, and the split and encoding of messages into common and private streams. We design the message split and the precoders of the radar sequence, common and private streams so as to jointly maximize the Weighted Sum Rate (WSR) and minimize the radar beampattern approximation Mean Square Error (MSE) subject to the per antenna power constraint. An iterative algorithm based on Alternating Direction Method of Multipliers (ADMM) is developed to solve the problem. 
Numerical results first show that RSMA-assisted DFRC achieves a better tradeoff between WSR and beampattern approximation than Space-Division Multiple Access (SDMA)-assisted DFRC with or without radar sequence, and other simpler radar-communication strategies using orthogonal resources. We also show that the RSMA-assisted DFRC frameworks with and without radar sequence achieve the same tradeoff performance. This is because that the common stream is better exploited in the proposed framework. The common stream of RSMA fulfils the triple function of managing interference among communication users, managing interference between communication and radar, and beampattern approximation. Uniquely, the SIC receiver of RSMA is exploited for the dual purpose of managing interference among communication users as well as interference between communication and radar. Therefore, by enabling RSMA in DFRC, the system performance is enhanced while the system architecture is simplified since there is no need to use additional radar sequence and SIC. We conclude that RSMA is a more powerful multiple access for DFRC. 
\end{abstract}
%

%
\IEEEpeerreviewmaketitle

\section{Introduction}
\subsection{Background and Literature Review}
In 5G and beyond, many promising applications, such as autonomous vehicles, Wi‐Fi sensing and extended reality have encouraged the Integration of Sensing and Communication (ISAC). As a typical and reliable means of sensing, radar has been widely used in both civil and military scenarios. Due to the scarcity of the spectrum resources, the 4th and 5th generation wireless communication systems are competing with radar applications in the S-band (2-4GHz), C-band (4-8GHz) and possibly millimeter-wave (mmWave) band, which may result in severe spectrum congestion and hamper the higher data rate demand in future wireless communications\cite{6967722}. As a long-term solution, communication and radar spectrum sharing raises wide interests and shows a significant potential for ISAC. There are two main research topics in the field of communication and radar spectrum sharing, namely, 1) coexistence of existing radar and communication devices, 2) Dual-Functional Radar-Communication (DFRC) system design \cite{LiuJointoverview}.\par 
For coexistence of multi-antenna radar and communication devices, the authors of \cite{Tang2019MIMOspec} design waveforms of Multiple-Input Multiple-Output (MIMO) radar to suppress the spectral energy leaked to the bands of communication devices, while the authors of \cite{Liu2017cellular} design the precoders for the Multi-User MIMO (MU-MIMO) communication to maximize the probing probability in a MIMO radar coexisting scenario and ensure the Signal-to-Interference-plus-Noise-Ratio (SINR) of downlink users. Instead of solely designing the multi-antenna radar device or communication devices, \cite{li2017joint} jointly designs the precoders of a MIMO radar and the codebook of a MIMO communication system to maximize radar effective SINR and ensure the average communication rate. \cite{qian2018joint} also jointly designs the transmit waveforms and receiver filters of a MIMO radar and the codebook of a MIMO communication system for the same purpose to achieve coexistence. 
However, in the literature of the coexistence of radar and communication devices, some important phenomena such as the cluster in radar detection are not considered in the simplified system models \cite{Zheng2019Overview}. Moreover, governmental and military agencies would be unwilling to make major changes in their radar deployments, which may impose huge costs on their financial budgets \cite{Liu2019HOW}. \par 
 Accounting for the possible drawbacks mentioned above, designing a DFRC system that makes the best use of the spectrum for both detecting and communicating is a promising alternative. Based on the waveform diversity of MIMO radar and the concept of Space-Division Multiple Access (SDMA) in MIMO communication, \cite{hassanien2015dual,hassanien2016phase} embed the information stream into radar pulses via a multi-antenna platform to detect targets at the mainlobe and transmit information streams at the sidelobe. The sidelobe level is modulated via Amplitude Shift Keying (ASK), where different power levels correspond to different communication symbols in \cite{hassanien2015dual}. The authors in \cite{hassanien2016phase} further develop Phase Shift Keying (PSK) in this system by representing the symbols as different phases of the signals received at the angle of the sidelobe. One significant restriction of such dual-functional system is that the communication rate is limited by the Pulse Repetition Frequency (PRF), which is far from satisfactory for communication requirements. To overcome this problem, \cite{liu2018mu} proposes a multi-antenna DFRC system that utilizes the communication streams for target detection, simultaneously transmits probing signals to radar targets and serves multiple downlink users. Specifically, the precoders are designed using the Zero-Forcing (ZF) beamforming technique to form a desired radar beampattern and meet the SINR requirements for communication users. Based on the transmission structure of \cite{liu2018mu}, the authors in \cite{xU2020Access} further investigate the tradeoff between Weighted Sum-Rate (WSR) and probing power at target for the multi-antenna DFRC, and show that a shared deployment with an integrated DFRC transmission outperforms the separated deployment (which is a special case of radar-communication coexistence). To the best of our knowledge, all existing works in the realm of multi-antenna DFRC consider the use of SDMA. Managing interference in DFRC can however be greatly improved by adopting more advanced multiple access techniques.\par  
Rate-Splitting Multiple Access (RSMA) has been recognized as a generalized and powerful physical-layer transmission framework for multi-antenna multi-user transmission networks that encompasses (and outperforms) SDMA and (power-domain) Non-Orthogonal Multiple Access (NOMA) as sub-schemes \cite{mao2018rate,clerckx2020WCL,mao2019RSMN,MAO2020BDPC}. As a building block of RSMA, Rate-Splitting (RS) splits the messages of users into private and common parts, jointly encodes the common parts into common streams to be decoded by multiple users and independently encodes the private parts into the private streams to be decoded by the corresponding users only, which enables to partially decode the interference and partially treat the remaining interference as noise. This significantly contrasts with SDMA that fully treats interference as noise and NOMA that fully decodes the interference. RSMA has been shown to achieve higher spectral and energy efficiencies than SDMA, NOMA, and Orthogonal Multiple Access (OMA) in multi-group multicast \cite{Joudeh2017MMB}, massive MIMO \cite{Dai2016MM}, millimeter-wave systems\cite{Dai2017MMwave}, Simultaneous Wireless Information and Power Transfer (SWIPT)\cite{mao2019RSWIPT}, etc. We are thus motivated to enable RSMA in DFRC to better manage the interference between radar and communication functions and also among communication users. \par 
\subsection{Contributions}
In this paper, we propose a RSMA-assisted multi-antenna DFRC system that simultaneously communicates with downlink users as a Base Station (BS) and probes targets with a desired beampattern as a collocated MIMO radar within the same frequency band. The major contributions are summarized as follows: \par 

\subsubsection{We propose a novel multi-antenna DFRC architecture built upon RSMA}
We exploit the powerful RSMA framework to further manage the interference between radar and communication functions as well as among communication users in multi-antenna DFRC. Specifically, the messages are split into the common and private parts and encoded respectively into the common stream, which is decoded by all users, and private streams, which are decoded by the corresponding users only. The common stream and private streams are precoded and transmit as dual-functional signals. By utilizing the common stream to partially decode the interference, RSMA achieves a significant advantage in DFRC to manage the interference between the dual functions and among communication users. To the best of our knowledge, this is the first work that enables RSMA in DFRC.
\subsubsection{We propose to enable the radar sequence apart from the communication streams in DFRC}
To improve the beampattern matching capability, we enable an additional radar sequence which is precoded and transmitted together with communication streams in the proposed multi-antenna DFRC. To the best of our knowledge, this is the first work to consider joint transmission of radar and communication signals in multi-antenna DFRC. We show that by using radar sequence, the tradeoff performance of conventional SDMA-assisted DFRC is improved. It is also shown that in RSMA-assisted DFRC, enabling and disabling radar sequence share the same dual-functional performance, which is explicitly much better than that of SDMA-assisted DFRC with radar sequence. This reveals that the common stream of RSMA not only has the same functions as the radar sequence in achieving desired beampattern and managing interference between radar and communication, but also has an advantage of managing the interference among communication users. Therefore, there is no need to use an additional radar sequence in RSMA-assisted DFRC. \par 
\subsubsection{We design an Alternating Direction Method of Multipliers (ADMM)-based optimization framework to solve the joint WSR maximization and Mean Square Error (MSE) of beampattern approximation minimization problem in DFRC}
The proposed DFRC is designed via solving a highly non-convex and complicated optimization problem whose objective function is a combination of maximizing the logarithmic WSR and minimizing the quartic MSE. In order to solve this problem, we propose an iterative framework based on ADMM that transforms the intractable problem into tractable iterative sub-problems. To solve the non-convex sub-problems in each iteration of ADMM, we devise an algorithm based on Majorization-Minimization (MM). 
We show that the proposed framework solves the non-convex optimization problem efficiently.\par 
\subsubsection{We present SDMA-assisted DFRC, Time-Division Radar-Communication (TDRC) and Frequency-Division Radar-Communication (FDRC) as baselines for a thorough comparison}  
First, to illustrate the advantages of integrated DFRC transmission using dual-functional signals, we introduce two practically simpler radar-communication strategies that operate the radar and communication functions via orthogonal resources in time and frequency domain, namely, TDRC and FDRC. Second, we develop the SDMA-assisted DFRC by turning off the common stream as baselines. The proposed DFRC has an advantage of being able to realizing dual functions within the same time-frequency resources compared with TDRC and FDRC. We show that the proposed RSMA-assisted DFRC outperforms SDMA-assisted DFRC as well as FDRC, and in certain conditions surpasses TDRC.     
\subsection{Organization}
The rest of the paper is organized as follows. The system model of the proposed DFRC is illustrated in Section II. In Section III, the metrics used are described, and the optimization problems are formulated. The ADMM-based iterative framework is specified in Section IV, and the algorithms for solving the updates in each ADMM iteration are illustrated in Section V. Section VI demonstrates the simulation results. Section VII concludes the paper.
\subsection{Notation}
Boldface upper-case and lower-case letters denote matrices and column vectors, respectively. The scalars are denoted by italic letters. Scalar $x_i$ denotes the $i$th entry of the vector $\bf x$. ${\bf I}_{N}$\ is the $N$-dimensional identity matrix, ${\bf 0}^{N\times K}$ is an all-zero matrix with dimension $N\times K$, and ${\bf 1}^{N\times K}$ is an all-one matrix with dimension $N\times K$. $(\cdot)^T$, $(\cdot)^H$, $(\cdot)^{-1}$ and $(\cdot)^{\dag}$ denote transpose, conjugate-transpose, inverse and pseudo inverse, respectively. $\lvert\cdot\lvert$ and $\lVert\cdot\lVert_2$ refer to the absolute value of a scalar and the 2-norm of a vector. The trace is $\text{tr}(\cdot)$, $\mathfrak{R}\{\cdot\}$ and $\mathfrak{J}\{\cdot\}$ are the real and conjugate part of a complex vector, respectively. Vector $\text{vec}({\bf X})$ is constructed by stacking the columns of ${\bf X}$. The vector $\text{diag}({\bf X})$ is the diagonal entries of ${\bf X}$. $\lambda_{\text{max}}({\bf X})$ and $\lambda_{\text{min}}({\bf X})$ respectively represent the largest and smallest eigenvalue of ${\bf X}$. Notation ${\bf A}\succeq {\bf B}$ stands for matrix ${\bf A}-{\bf B}$ is positive semidefinite.

%
%
%
%

\section{System Model}
In this work, we focus on a novel multi-antenna DFRC that supports ISAC. Specifically, the transmitter is equipped with $N_{\text{t}}$ antennas, aiming at serving $K$ single-antenna downlink communication users as a BS and detecting multiple radar targets in the azimuth angles of interests as a collocated MIMO radar. For simplicity of exposure, we assume the transmitter adopts a Uniform Linear Array (ULA). The communication users are indexed by $\mathcal{K} =\{1,\dots, K\}$. In the following subsections, we will specify the proposed DFRC system model with RSMA and radar sequence enabled, followed by the three baseline schemes, namely, SDMA-assisted DFRC, TDRC, FDRC.\par

\subsection{Proposed RSMA-assisted DFRC}\label{ProposeRSMAsec}
\begin{figure*}[htb]
	\centering
	\includegraphics[width=0.75\linewidth]{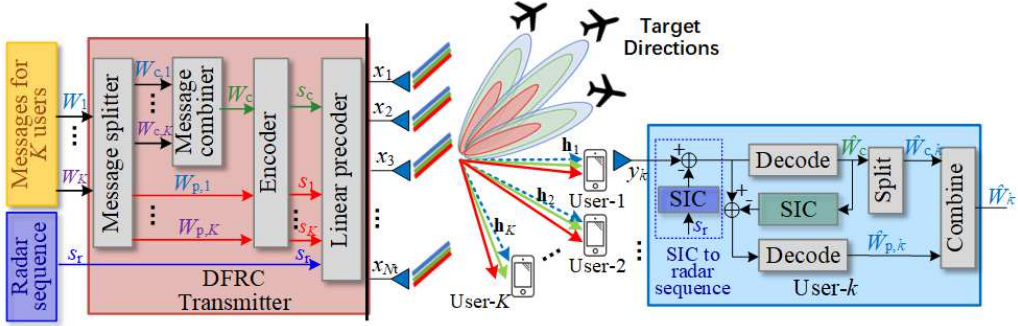}
	\caption{System architecture of the proposed RSMA-assisted DFRC}\label{rsframe}
\end{figure*}
As is shown in Fig. \ref{rsframe}, the proposed DFRC enables the simultaneous transmission of the precoded information streams and the radar sequence to fulfil both radar detection and communication functions. RSMA proposed in \cite{mao2018rate} is enabled at the transmitter to process the messages intended for the communication users. Specifically, the message $W_k$ of the $k$th user is split into a common part $W_{\text{c},k}$ and a private part $W_{\text{p},k}$, $\forall k\in \mathcal{K}$. The common parts of all users $\{W_{c,1}, \dots, W_{\text{c},k}\}$ are jointly encoded into the common stream $s_{\text{c}}$, while the private parts $\{W_{p,1}, \dots, W_{\text{p},k}\}$ are respectively encoded into private streams $\{s_{1}, \dots, s_{K}\}$. Then the data streams as well as the transmit radar sequence, which are denoted together as ${\bf s}=[s_{\text{c}},s_1,\dots,s_k,s_{\text{r}}]^T \in \mathbb{C}^{K+2}$, are linearly precoded before transmission using the precoder ${\bf P}=[{\bf{p}}_{\text{c}},{\bf p}_1,\dots,{\bf{p}}_K,{\bf{p}}_{\text{r}}]$, where ${\bf{p}}_{\text{c}}\in \mathbb{C}^{N_t}$ is the precoder of the common stream and ${\bf p}_{\text{r}}\in \mathbb{C}^{N_{\text{t}}}$ is the precoder of the radar sequence. We denote the power of the common stream as $P_{\text{c}}=\lVert{\bf p}_{\text{c}}\lVert_2$ and that of the radar sequence as $P_{\text{r}}=\lVert{\bf p}_{\text{r}}\lVert_2$. The baseband transmit signal can be fully expressed as 
\begin{equation}
{\bf x}=\underbrace{\overbrace{{\bf p}_{\text{c}}s_{\text{c}}}^{\text{Common stream}}+\overbrace{\sum_{k\in \mathcal{K}}{\bf p}_k{s}_k}^{\text{Private streams}}}_{\text{Communication streams}}+\underbrace{{\bf p}_{\text{r}}s_{\text{r}}}_{\text{Radar sequence}}.\label{CRtransmit}
\end{equation}
Note that the transmit radar sequence $s_{\text{r}}$ contains no useful messages for users and thus is not random. To improve detection performance, the waveform of the radar sequence is usually designed and determined previously before transmission. We set aside the radar sequence waveform design problem as future work and focus on the precoder design. We assume that ${\bf s}$ follows $\mathbb{E}\{{\bf s}{\bf s}^H\}={\bf I}_{(K+2)}$. The total transmit power budget is $P_\text{t}$, i.e., $\text{tr}({\bf P}{\bf P}^H)\leq P_\text{t}$. \par 

Based on \eqref{CRtransmit}, the received signal at the $k$th downlink communication user is
\begin{equation}
\begin{aligned}
y_k=&{\bf h}_k^H{\bf x}+n_k\\
=&{\bf h}_k^H{\bf p}_{\text{r}}s_{\text{r}}+{\bf h}_k^H{\bf p}_{\text{c}}s_{\text{c}}+{\bf h}_k^H\sum_{j\in \mathcal{K}}{\bf p}_j{s}_j+n_k
\end{aligned}\label{CRreceive}
\end{equation} 
where ${\bf h}_k \in \mathbb{C}^{{N_{\text{t}}}\times 1}$ is the channel vector between the DFRC transmitter and the $k$th user, which is assumed to be perfectly known at the transmitter and the users. Received noise $n_k$ at the $k$th user is modelled as a complex Gaussian random variable with zero mean and variance $\sigma _{n,k}^2$. Without loss of generality, we assume the noise variances of all users equal to 1, i.e., $\sigma _{n,k}^2=1, \forall k \in \mathcal{K}$. \par 
Depending on whether to utilize SIC to cancel the radar sequence interference at users, there are different radar sequence modes of the proposed DFRC, which lead to different metric expressions in the subsequent problem formulation.
\subsection{Different Modes of Radar Sequence}\label{differentradarmodes}
By introducing $\delta_{\text{c}}$ as an index variable, all three possible modes of the radar sequence are specified in this part.
\subsubsection{Radar sequence is disabled} In this mode, the radar sequence is completely turned off in the proposed DFRC transmission framework. This is equivalent to allocating zero power to ${\bf p}_{\text{r}}$ in all above expressions, i.e., $P_{\text{r}}=0$. In the following metrics and problem formulation, we specify $\delta_c=0$ and $P_{\text{r}}=0$ to represent this mode. We define this mode as DFRC without radar sequence in the following parts.
\subsubsection{Radar sequence is enabled with SIC} The DFRC system transmits the radar sequence in this mode. However, since the radar sequence does not contain messages, we assume that $s_{\text{r}}$ is pre-stored and known to the users. Therefore, its interference is successfully canceled via SIC before information streams are decoded. We let $\delta_{\text{c}}=0$ to denote this mode and define it as DFRC with radar sequence and SIC.
\subsubsection{Radar sequence is enabled without SIC} For easier implementation, SIC is not applied to $s_{\text{r}}$, which means the radar sequence will be treated as interference at users when information streams are decoded. We use $\delta_{\text{c}}=1$ to denote this mode in the following parts and define it as DFRC with radar sequence.\par
\subsection{SDMA-assisted DFRC Baselines} 
In order to show the advantages of RSMA, we also study DFRC enabling SDMA based on Multi-User Linear Precoding (MU-LP) \cite{mao2018rate} as a baseline. SDMA-assisted DFRC can be realised by disabling the common stream in \eqref{CRtransmit} and \eqref{CRreceive}, i.e., letting $P_{\text{c}}=0$. However, in the proposed DFRC system, switching between RSMA and SDMA does not influence the working mode of the radar sequence.\par 
To be clear, we list variable settings for all possible radar sequence modes and multiple access methods in Table \ref{varset}.  \par 
\begin{table}[t]
	\centering
	\caption{Variable Settings For Different Combinations of Multiple Access Techniques and Radar Sequence Modes in the Proposed DFRC}  
	\begin{tabular}{ccc}
		\toprule
		& RSMA-assisted DFRC & SDMA-assisted DFRC \\ \hline
		\begin{tabular}[c]{@{}c@{}}Radar sequence\\ disabled  \end{tabular}                            & 	$P_{\text{r}}=0$, $\delta_{\text{c}}=0$    & \begin{tabular}[c]{@{}c@{}}$P_{\text{c}}=0$, $P_{\text{r}}=0$,\\ $\delta_{\text{c}}=0$\end{tabular}     \\ \hline
		\begin{tabular}[c]{@{}c@{}} Radar sequence \\enabled with SIC\end{tabular}    &  $\delta_{\text{c}}=0$    &  $P_{\text{c}}=0$, $\delta_{\text{c}}=0$    \\ \hline
		\begin{tabular}[c]{@{}c@{}} Radar sequence\\ enabled without SIC\end{tabular} & $\delta_{\text{c}}=1$     & $P_{\text{c}}=0$, $\delta_{\text{c}}=1$      \\ 	\bottomrule
	\end{tabular}\label{varset}
\end{table}
\subsection{TDRC and FDRC Baselines}
To show the advantages of the proposed DFRC transmission architecture, we compare it with practically simpler radar-communication strategies using orthogonal resources in time and frequency domain, i.e., TDRC and FDRC. Both strategies are specified as follows. We denote the power budget of radar and communication functions as $P_{\text{R}}, P_{\text{C}}$, respectively.
\begin{table}[t]
	\centering
	\caption{Resources Used in the Proposed DFRC, TDRC and FDRC }  
	\begin{tabular}{@{}ccccc@{}}
		\toprule
		\multirow{2}{*}{} & \multicolumn{2}{c}{Time Allocation} & \multirow{2}{*}{\begin{tabular}[c]{@{}c@{}}Within the Same \\ Frequency Band\end{tabular}} & \multirow{2}{*}{\begin{tabular}[c]{@{}c@{}}Power Budget for \\ Each Function\end{tabular}} \\ \cmidrule(lr){2-3}
		& Comms           & Radar             &                                                                                            &                                                                                            \\ \midrule
		Proposed DFRC     & 1               & 1                 & Yes                                                                                        & $P_{\text{R}}=P_{\text{C}}=P_{\text{t}}$                                                   \\
		TDRC              & $\alpha$        & $1-\alpha$        & Yes                                                                                        & $P_{\text{R}}=P_{\text{C}}=P_{\text{t}}$                                                   \\
		FDRC              & 1               & 1                 & No                                                                                         & $P_{\text{R}}+P_{\text{C}}=P_{\text{t}}$                                                   \\ \bottomrule
	\end{tabular}\label{resourse}
\end{table}
\subsubsection{Time-Division Radar-Communication} The transmitter of the TDRC baseline spends a fraction $\alpha$ of time purely working as a $N_{\text{t}}$-antenna BS and $1-\alpha$ of time working as a $N_{\text{t}}$-antenna MIMO radar within the same frequency band. The working power budget is $P_{\text{t}}$ for either function duration. To ensure the fairness between TDRC and the proposed RSMA-assisted DFRC, we assume that RSMA is enabled when the transmitter acts as a BS. The value of $\alpha$ depends on real task demands. Since different $\alpha$ leads to different tradeoffs, we only adopt representative values of $\alpha$ as baselines in numerical experiments for an explicit comparison. 
\subsubsection{Frequency-Division Radar-Communication} The transmitter of the FDRC simultaneously transmits precoded communication streams and radar probing signals within different frequency bands. Since FDRC operates radar and communication functions both for the whole time duration, the total power budget is split for both functions, i.e.,  $P_{\text{R}}+P_{\text{C}}=P_{\text{t}}$. There is thus no interference between the radar and communication functions because of the frequency orthogonality. RSMA is enabled for the transmission of the communication streams.\par 
The resource allocation for the proposed DFRC, TDRC and FDRC are compared in Table \ref{resourse}. It is obvious that DFRC has an advantage of simultaneously fulfilling dual functions within the same frequency band.

\section{Metrics and problem formulation}
In this section, we introduce two classic metrics, i.e., WSR for communication and MSE of beampattern approximation for radar. These two metrics are used to formulate the optimization problem of DFRC transmission design.
\subsection{Metric for Communication: WSR}
To evaluate the communication performance of DFRC, we use WSR in our work, since this is the most representative performance metric of modern multiuser communication systems. \par 
According to the RSMA-assisted DFRC strategy proposed in Subsection \ref{ProposeRSMAsec} and different modes of radar sequence defined in Subsection \ref{differentradarmodes}, the respective SINRs of decoding $s_{\text{c}}$ and $s_{k}$ at the $k$th user are
\begin{align}
\gamma_{\text{c},k}({\bf P},\delta_{\text{c}})=&\frac{\left| {\bf h}_k^H{\bf p}_{\text{c}}\right| ^2}{\sum_{j\in \mathcal{K}}\left| {\bf h}_k^H{\bf p}_j\right|^2+\delta_{\text{c}}\left| {\bf h}_k^H{\bf p}_{\text{r}}\right|^2+1},\forall k \in \mathcal{K},\label{rck_CR-noSIC}\\
\gamma_k({\bf P},\delta_{\text{c}})=&\frac{\left| {\bf h}_k^H{\bf p}_k\right| ^2}{\sum_{j\in \mathcal{K},j\neq k}\left| {\bf h}_k^H{\bf p}_j\right|^2+\delta_{\text{c}}\left| {\bf h}_k^H{\bf p}_{\text{r}}\right|^2+1},\forall k \in \mathcal{K},\label{rk_CR-noSIC}
\end{align}
where the index variable $\delta_{\text{c}}$ is used to unify the expressions for different radar sequence processing modes. Specifically, when the radar sequence is enabled and is not removed via SIC, i.e., $\delta_{\text{c}}=1$, it is fully treated as interference when each user-$k$ decodes $s_\text{c}$ and $s_k$. Otherwise, when $\delta_c=0$, it is fully removed from the received signal at each user. Based on \eqref{rck_CR-noSIC} and \eqref{rk_CR-noSIC}, the achievable rates of $s_{\text{c}}$ and $s_k$ at the $k$th user are
\begin{align}
R_{\text{c},k}({\bf P},\delta_{\text{c}})=&\log_2\left(1+\gamma_{\text{c},k}({\bf P},\delta_{\text{c}})\right),\\
R_{k}({\bf P},\delta_{\text{c}})=&\log_2\left(1+\gamma_{k}({\bf P},\delta_{\text{c}})\right).
\end{align}\par 
To ensure that the common stream is successfully
decoded by all users, the rate of decoding the common stream $s_{\text{c}}$ should not exceed $R_{\text{c}}({\bf P},\delta_{\text{c}})=\min \left\{R_{\text{c},1}({\bf P},\delta_{\text{c}}),\dots,R_{\text{c},k}({\bf P},\delta_{\text{c}})\right\}$\cite{mao2018rate}. 
As $R_{\text{c}}({\bf P},\delta_{\text{c}})$ is shared by $K$ users, we have $\sum_{k\in\mathcal{K}}C_k=R_{\text{c}}({\bf P},\delta_{\text{c}})$,
where $C_k$ is the portion of common rate at user-$k$ transmitting $W_{\text{c},k}$. The total achievable rate of user-$k$ contains the portion of common rate transmitting $W_{\text{c},k}$ and the private rate transmitting $W_{\text{p},k}$. Thus, given the rate weight allocated to user-$k$ as $\mu_k$, the WSR for the proposed RSMA-assisted DFRC system is 
\begin{equation}
	{\text{WSR}({\bf P},\delta_{\text{c}})}=\sum_{k\in \mathcal{K}}\mu_k\left(C_k+R_{k}({\bf P},\delta_{\text{c}})\right).\label{WSRexpression}
\end{equation}\par 
For SDMA-assisted DFRC, the corresponding WSR expression is obtained from (7) by turning off $C_k,
\forall k\in\mathcal{K}$, which is the WSR of the private streams only.

\subsection{Metric for Radar: MSE of Beampattern Approximation }
It is well-known that to get higher Signal-to-Noise-Ratio (SNR) for target tracking or narrow-beam scanning, a specially desired rather than a nominally omnidirectional MIMO radar beampattern is expected to be achieved via correlated waveform design \cite{fuhrmann2008transmit,stoica2007probing}.
Hence, we focus on the transmit beampattern design to represent radar performance in this work. According to existing literature \cite{ fuhrmann2008transmit,stoica2007probing}, beampattern design is generally formulated into a desired beampattern approximation problem via least squares. Based on the metric of these works, we introduce MSE for beampattern approximation as
$
\sum_{m=1}^{M}\left| P_{\text{d}}(\theta_m)-{\bf a}^H(\theta_m){\bf R}_{\bf x}{\bf a}(\theta_m)\right|^2,
$ 
where $\theta_m$ is the $m$th azimuth angle grid among all $M$ grids, $P_{\text{d}}(\theta_m)$ is the desired beampattern level at $\theta_m$, $\bf R_x$ is the covariance matrix of transmit waveforms, ${\bf a}(\theta_m)=[1,e^{\text{j}2\pi\Delta \sin(\theta_m)},\dots,e^{\text{j}2\pi(N_{\text{t}}-1)\Delta \sin(\theta_m)}]^T\in\mathbb{C}^{N_{\text{t}}\times 1}$ is the transmit steering vector for ULA and $\Delta$ is the normalized distance (relative to wavelength) between adjacent array elements. To improve robustness, $P_{\text{d}}(\theta_m)$ is assumed to be generated previously via classical beampattern synthesis approaches such as in \cite{stoica2007probing} according to the radar prior knowledge including azimuth angles of interest and the array setting.\par 
For the transmission of our proposed DFRC system, we thus have the beampattern approximation MSE as
\begin{equation}
\text{MSE}({\bf P})=\sum_{m=1}^{M}\left| P_{\text{d}}(\theta_m)-{\bf a}^H(\theta_m){\bf P}{\bf P}^H{\bf a}(\theta_m)\right|^2.\label{MSE}
\end{equation}
When $P_{\text{c}}=0$, \eqref{MSE} is the MSE for SDMA-assisted DFRC. Likewise, by forcing $P_{\text{r}}=0$, \eqref{MSE} is the MSE when radar sequence is disabled.

\subsection{Problem Formulation}
In this work, we focus on jointly designing the message split, precoders of communication streams and radar sequence with the aim of jointly maximizing WSR for communication users and minimizing MSE for radar beampattern approximation. To keep the tradeoff between WSR and MSE, a regularization parameter $\lambda\in[0,1]$ is introduced to combine WSR and MSE in the objective function. The optimization problem of our proposed RSMA-assisted DFRC strategy is formulated as
	\begin{subequations}\label{CR origin}
	\begin{align}
	\max\limits_{{\bf c},{\bf P}} \quad &(1-\lambda)\sum_{k\in \mathcal{K}}\mu_k\left(C_k+R_{k}({\bf P},\delta_{\text{c}})\right)\notag\\&-\lambda\sum_{m=1}^{M}\left| P_{\text{d}}(\theta_m)-{\bf a}^H(\theta_m){\bf P}{\bf P}^H{\bf a}(\theta_m)\right|^2\label{CR origin c3} \\ 
	s.t.\quad&\sum_{k'\in \mathcal{K}}C_{k'}\leq R_{\text{c},k}({\bf P},\delta_{\text{c}}),\forall k\in\mathcal{K}\label{CR origin c1}\\
	&{\bf c}\geq 0\label{CR origin c2}\\
	&\text{diag}({\bf P}{\bf P}^H)=\frac{P_{\text{t}}{\bf 1}^{N_{\text{t}}\times 1}}{N_{\text{t}}},\label{CR origin c4}
	\end{align}
	\end{subequations}
where ${\bf c}=[C_{1},C_2,\dots,C_K]^T$. \eqref{CR origin c1} ensures the common stream is successfully decoded at each user. \eqref{CR origin c4} ensures the transmit average power of each antenna to be the same, which is commonly adopted for MIMO radar beampattern design problem \cite{fuhrmann2008transmit,stoica2007probing}. \par 
Note that by applying the variable settings in Table \ref{varset} to \eqref{CR origin}, we obtain the optimization problems of SDMA-assisted baseline schemes with different radar sequence modes. 

\section{ADMM-based Framework for Solving the Problem }\label{admmdetail}
The combination of logarithmic WSR and quartic MSE of beampattern approximation in the objective function is non-convex. The equality power constraint is also non-convex and intractable. Since \eqref{CR origin} combines both the objective of maximizing the WSR for communication users and minimizing the MSE for radar beampattern approximation, our idea is to decompose the problem into two communication and radar sub-problems, which are then solved alternatively until a local optimal solution is obtained. \par 
To give an explicit expression of the framework, we first denote a new vector that contains all optimization variables in \eqref{CR origin}, i.e., ${\bf v}=[{\bf c}^T,{\bf p}_{\text{vec}}^T]^T\in \mathbb{R}_+^{K}\times\mathbb{C}^{N_{\text{t}}(K+2)}$, with ${\bf p}_{\text{vec}}=\text{vec}({\bf P})$. 
We then define selection matrices as
$
{\bf D}_{\text{p}}=\left[{\bf 0}^{(K+2)N_{\text{t}}\times K}, {\bf I}_{(K+2)N_{\text{t}}}\right]$, $
{\bf D}_{\text{c}}=\left[{\bf 0}^{N_{\text{t}}\times K}, {\bf I}_{N_{\text{t}}}, {\bf 0}^{N_{\text{t}}\times (K+1)N_{\text{t}}}\right]$, $
{\bf D}_{\text{r}}=\left[{\bf 0}^{N_{\text{t}}\times [K+(K+1)N_{\text{t}}]}, {\bf I}_{N_{\text{t}}}\right]$, $
{\bf D}_k=\left[{\bf 0}^{N_{\text{t}}\times (K+kN_{\text{t}})}, {\bf I}_{N_{\text{t}}}\quad {\bf 0}^{N_{\text{t}}\times (K+1-k)N_{\text{t}}}\right]
$. We also define selection vectors $
{\bf e}_k=[{\bf 0}^{1\times(k-1)}, 1, {\bf 0}^{1\times[(K+2)N_\text{t}+K-k]}]^T$, such that we have $C_k={\bf e}_k^T{\bf v}$.\par
To unify the expressions, when the radar sequence mode is determined, we denote the rates at user-$k$ as 
$
\eta^{\delta_{\text{c}}}_{\text{c},k}({\bf p}_{\text{vec}})=\eta^{\delta_{\text{c}}}_{\text{c},k}({\bf D}_{\text{p}}{\bf v})=R_{\text{c},k}({\bf P},\delta_{\text{c}})$, $\eta^{\delta_{\text{c}}}_{k}({\bf p}_{\text{vec}})=\eta^{\delta_{\text{c}}}_{k}({\bf D}_{\text{p}}{\bf v})=R_{k}({\bf P},\delta_{\text{c}}),
$
and rewrite \eqref{CR origin} into a tractable ADMM formulation as
\begin{equation}
\begin{split}
\min\limits_{{\bf v},{\bf u}} \quad &f_{\text{c}}({\bf v})+g_{\text{c}}({\bf v})+f_{\text{r}}({\bf u})+g_{\text{r}}({\bf u})\\ 
s.t.\quad &{\bf D}_{\text{p}}({\bf v}-{\bf u})=0,
\end{split} \label{CO admm}
\end{equation}
where $
f_{\text{c}}({\bf v})=-(1-\lambda)\sum_{k\in \mathcal{K}}\mu_k\left({\bf e}_k^T{\bf v}+\eta^{\delta_{\text{c}}}_{k}({\bf D}_{\text{p}}{\bf v})\right)$. $g_{\text{c}}({\bf v})$ is the indicator function of the communication feasible set $\mathcal{C}=\left\{{\bf v}\bigg|\sum_{k\in \mathcal{K}}{\bf e}_k^T{\bf v}\leq \eta^{\delta_{\text{c}}}_{\text{c},k}({\bf D}_{\text{p}}{\bf v}), {\bf v}^H{\bf D}_{\text{p}}^H{\bf D}_{\text{p}}{\bf v}\leq P_{\text{t}}\right\}$, where the first condition corresponds to \eqref{CR origin c1}. The second condition is added as a power constraint only to improve robustness of convergence, which does not change the original problem as it is always met if \eqref{CR origin c4} holds. ${\bf u}\in \mathbb{R}_+^{K}\times\mathbb{C}^{N_{\text{t}}(K+2)}$ is introduced as a new vector variable.
$ 
 f_{\text{r}}({\bf u})=\lambda\sum_{m=1}^{M}|P_{\text{d}}(\theta_m)-{\bf a}^H(\theta_m)\big({\bf D}_{\text{c}}{\bf u}{{\bf u}}^H{\bf D}_{\text{c}}^H+\sum_{k\in \mathcal{K}}{\bf D}_k{\bf u}{{\bf u}}^H{\bf D}_k^H+{\bf D}_{\text{r}}{\bf u}{{\bf u}}^H{\bf D}_{\text{r}}^H \big) {\bf a}(\theta_m)| ^2
$, and 
 $g_{\text{r}}({\bf u})$ is the indicator function of radar feasible set $
 \mathcal{R}=\left\{{\bf u}\bigg|\text{diag}({\bf D}_{\text{c}}{\bf u}{{\bf u}}^H{\bf D}_{\text{c}}^H+\sum_{k\in \mathcal{K}}{\bf D}_k{\bf u}{{\bf u}}^H{\bf D}_k^H\right.$ $\left.+{\bf D}_{\text{r}}{\bf u}{{\bf u}}^H{\bf D}_{\text{r}}^H)=\frac{P_{\text{t}}{\bf 1}^{N_{\text{t}}\times 1}}{N_{\text{t}}}\right\}$ corresponding to \eqref{CR origin c4}.  \par 
According to \cite{MAL-016}, \eqref{CO admm} can be solved by iteratively updating ${\bf v}, {\bf u}, {\bf d}$, which is given as
\begin{align}
{\bf v}^{t+1}:=&\arg\min\limits_{{\bf v}}\big(f_{\text{c}}({\bf v})+g_{\text{c}}({\bf v})\notag\\
&+(\rho/2)\lVert {\bf D}_{\text{p}}({\bf v}-{\bf u}^t)+{\bf d}^t\rVert_2^2\big),\label{updatev}\\
{\bf u}^{t+1}:=&\arg\min\limits_{{\bf u}}\big(f_{\text{r}}({\bf u})+g_{\text{r}}({\bf u})\notag\\
&+(\rho/2)\lVert{\bf D}_{\text{p}}({\bf v}^{t+1}-{\bf u})+{\bf d}^t\rVert_2^2\big),\label{updateu}\\
{\bf d}^{t+1}:=&{\bf d}^t+ {\bf D}_{\text{p}}({\bf v}^{t+1}-{\bf u}^{t+1}),\label{updated}
\end{align}
where ${\bf d}\in \mathbb{C}^{N_{\text{t}}(K+2)}$ is the scaled dual variable. It is worth clarifying that \eqref{updatev} -- \eqref{updated} can be rewritten into real-value formations following the rules in \cite{Luo2010Semidefinite} in order to rigorously meet the real-value definition of ADMM \cite{MAL-016}.\par
The proposed ADMM-based framework is summarized in Algorithm 1, where ${\bf r}^{t+1}$ and ${\bf q}^{t+1}$ are the primal and dual residuals, and $\epsilon_0$ is the feasibility tolerance. Based on the proposed framework, we are able to solve the intractable problem \eqref{CR origin} through updating ${\bf v}, {\bf u}, {\bf d}$, respectively. \par 
The update of $\bf{v}$ ($\bf v$-update) in \eqref{updatev} is achieved by using the Weighted Minimum Mean Square Error (WMMSE) algorithm while the update of $\bf{u}$ ($\bf u$-update) in \eqref{updateu} is based on MM algorithm. Both algorithms will be specified in Section \ref{algorithmsforupdates} followed by the computational complexity analysis of the entire ADMM framework.
\begin{algorithm}\label{ADMMAL}
	\caption{ADMM-based Framework}
	\LinesNumbered 
	\KwIn{$t\leftarrow0,{\bf v}^{t},{\bf u}^{t},{\bf d}^{t}$}
	\Repeat{$\lVert{\bf r}^{t+1}\lVert_2\le \epsilon_0$ and $\lVert{\bf q}^{t+1}\lVert_2\le \epsilon_0$}{
		Update ${\bf v}^{t+1}\gets \arg\min\limits_{{\bf v}}\big(f_{\text{c}}({\bf v})+g_{\text{c}}({\bf v})+(\rho/2)\lVert {\bf D}_{\text{p}}({\bf v}-{\bf u}^{t})+{\bf d}^{t}\rVert_2^2\big)$ via WMMSE algorithm;\\
		Update ${\bf u}^{t+1}\gets \arg\min\limits_{{\bf u}}\big(f_{\text{r}}({\bf u})+g_{\text{r}}({\bf u})+(\rho/2)\lVert{\bf D}_{\text{p}}({\bf v}^{t+1}-{\bf u})+{\bf d}^{t}\rVert_2^2\big)$ via MM-based algorithm;\\
		Update ${\bf d}^{t+1}\gets {\bf d}^t+ {\bf D}_{\text{p}}({\bf v}^{t+1}-{\bf u}^{t+1})$;\\
		${\bf r}^{t+1}={\bf D}_{\text{p}}({\bf v}^{t+1}-{\bf u}^{t+1})$;\\
		${\bf q}^{t+1}={\bf D}_{\text{p}}({\bf u}^{t+1}-{\bf u}^{t})$;\\
		t++;
	}
\end{algorithm}

\section{Algorithms for solving ADMM updates}\label{algorithmsforupdates}
The ${\bf v}$-update and ${\bf u}$-update in the ADMM framework are still challenging. Thus, we adopt the WMMSE algorithm for ${\bf v}$-update and MM-based algorithm for ${\bf u}$-update in this section.
\subsection{WMMSE Algorithm for $\bf v$-update}
In this subsection, we illustrate the procedure of solving the ${\bf v}$-update in each ADMM iteration. Since ${\bf v}=[{\bf c}^T,{\bf p}_{\text{vec}} ^T]^T$, \eqref{updatev} is equivalently rewritten in a tractable manner as
\begin{equation}
\begin{split}\label{updatevorigin}
\min\limits_{{\bf c},{\bf p}_{\text{vec}}} \quad &-(1-\lambda)\sum_{k\in \mathcal{K}}\mu_k\left(C_k+\eta^{\delta_{\text{c}}}_k({\bf p}_{\text{vec}})\right)\\
&+\frac{\rho}{2}\lVert {\bf p}_{\text{vec}}-{\bf D}_{\text{p}}{\bf u}^t+{\bf d}^t\rVert_2^2\\ 
s.t.\quad&\sum_{k'\in \mathcal{K}}C_{k'}\leq \eta^{\delta_{\text{c}}}_{\text{c},k}({\bf p}_{\text{vec}}),\forall k\in\mathcal{K}\\
&{\bf p}_{\text{vec}}^H{\bf p}_{\text{vec}}\leq P_{\text{t}}\\
&{\bf c}\geq 0.\\
\end{split}
\end{equation}\par 
This problem can be reformulated with WMMSE approach and solved through the WMMSE algorithm following \cite{mao2019RSWIPT}.\par 
The mode index $\delta_{\text{c}}$ is used to unify the derivation for different radar sequence modes. The common stream $s_{\text{c}}$ is first decoded at user-$k$ via an equalizer $g^{\delta_{\text{c}}}_{\text{c},k}$. Equalizer $g^{\delta_{\text{c}}}_{k}$ is employed to decode the private stream $s_k$ after removing the decoded common stream. For both modes, we get the estimations $\hat{s}^{\delta_{\text{c}}}_{\text{c}}$, $\hat{s}^{\delta_{\text{c}}}_k$ of $s_{\text{c}}$, $s_k$ , which are respectively given as$
\hat{s}^{\delta_{\text{c}}}_{\text{c}}=g^{\delta_{\text{c}}}_{k}\tilde y^{\delta_{\text{c}}}_k$,
$\hat{s}^{\delta_{\text{c}}}_k=g^{\delta_{\text{c}}}_{k}(\tilde y^{\delta_{\text{c}}}_k-{\bf h}_k^H{\bf p}_{\text{c}}s_{\text{c}})$, where
\begin{equation}
	\tilde y^{\delta_{\text{c}}}_k=\left\{
\begin{array}{ll}
y_k-{\bf h}_k^H{\bf p}_{\text{r}}s_{\text{r}}, &\delta_{\text{c}}=0\\
y_k, &\delta_{\text{c}}=1
\end{array}
\right..
\end{equation} 
Subsequently, the MSEs of both common and private stream estimation, defined respectively as $\varepsilon^{\delta_{\text{c}}}_{\text{c},k}\triangleq\mathbb{E}\left\{\left| \hat{s}^{\delta_{\text{c}}}_{\text{c}}-s_{\text{c}}\right|^2\right\}$ and $\varepsilon^{\delta_{\text{c}}}_{k}\triangleq\mathbb{E}\left\{\left| \hat{s}^{\delta_{\text{c}}}_k-s_k\right|^2\right\}$, are expressed as
\begin{align}
	\varepsilon^{\delta_{\text{c}}}_{\text{c},k}&=\left|g_{\text{c},k}\right|^2T^{\delta_{\text{c}}}_{\text{c},k}-2\mathfrak{R}\left\{g^{\delta_{\text{c}}}_{\text{c},k}{\bf h}_k^H{\bf p}_{\text{c}}\right\}+1,\\
	\varepsilon^{\delta_{\text{c}}}_{k}&=\left|g_{k}\right|^2T^{\delta_{\text{c}}}_{k}-2\mathfrak{R}\left\{g^{\delta_{\text{c}}}_{k}{\bf h}_k^H{\bf p}_k\right\}+1,
\end{align}

where 
\begin{align}
&T^{\delta_{\text{c}}}_{\text{c},k}\triangleq \notag\\
&\left\{
\begin{array}{ll}
\left|{\bf h}_k^H{\bf p}_{\text{c}}\right|^2+\sum_{j\in \mathcal{K}}\left|{\bf h}_k^H{\bf p}_j\right|^2+1, &\delta_{\text{c}}=0\\
\left|{\bf h}_k^H{\bf p}_{\text{r}}\right|^2+\left|{\bf h}_k^H{\bf p}_{\text{c}}\right|^2+\sum_{j\in \mathcal{K}}\left|{\bf h}_k^H{\bf p}_j\right|^2+1, &\delta_{\text{c}}=1
\end{array}
\right.,\\
&T^{\delta_{\text{c}}}_{k}\triangleq\left\{
\begin{array}{ll}
\sum_{j\in \mathcal{K}}\left|{\bf h}_k^H{\bf p}_j\right|^2+1, &\delta_{\text{c}}=0\\
\left|{\bf h}_k^H{\bf p}_{\text{r}}\right|^2+\sum_{j\in \mathcal{K}}\left|{\bf h}_k^H{\bf p}_j\right|^2+1, &\delta_{\text{c}}=1
\end{array}
\right..
\end{align}
Optimal equalizers are obtained by letting $\frac{\partial\varepsilon^{\delta_{\text{c}}}_{\text{c},k}}{\partial g^{\delta_{\text{c}}}_{\text{c},k}}=0$ and $\frac{\partial\varepsilon^{\delta_{\text{c}}}_{k}}{\partial g^{\delta_{\text{c}}}_{k}}=0$, which are the Minimized MSE (MMSE) equalizers given by 
$
g^{\text{MMSE},{\delta_{\text{c}}}}_{\text{c},k}={\bf p}_{\text{c}}^H{\bf h}_k(T^{\delta_{\text{c}}}_{\text{c},k})^{-1}$, $
g^{\text{MMSE},{\delta_{\text{c}}}}_{k}={\bf p}_k^H{\bf h}_k(T^{\delta_{\text{c}}}_{k})^{-1}.
$
The MMSEs based on $g^{\text{MMSE},{\delta_{\text{c}}}}_{\text{c},k}$ and $g^{\text{MMSE},{\delta_{\text{c}}}}_{k}$ are given by
\begin{equation}
\begin{split}
	\varepsilon^{\text{MMSE},{\delta_{\text{c}}}}_{\text{c},k}&\triangleq \min\limits_{g^{\delta_{\text{c}}}_{\text{c},k}}\varepsilon^{\delta_{\text{c}}}_{\text{c},k}=(T^{\delta_{\text{c}}}_{\text{c},k})^{-1}\left(T^{\delta_{\text{c}}}_{\text{c},k}-\left|{\bf h}_k^H{\bf p}_{\text{c}}\right|^2\right),\\
	\varepsilon^{\text{MMSE},{\delta_{\text{c}}}}_{k}&\triangleq \min\limits_{g^{\delta_{\text{c}}}_{k}}\varepsilon^{\delta_{\text{c}}}_{k}=(T^{\delta_{\text{c}}}_{k})^{-1}\left(T^{\delta_{\text{c}}}_{k}-\left|{\bf h}_k^H{\bf p}_k\right|^2\right).
\end{split}\label{MMSEs}
\end{equation}\par 
Hence, by comparing \eqref{MMSEs} with \eqref{rck_CR-noSIC} and \eqref{rk_CR-noSIC}, we rewrite SINRs of decoding the intended streams as $\gamma^{\delta_{\text{c}}}_{\text{c},k}=(1/\varepsilon_{\text{c},k}^{\text{MMSE},{\delta_{\text{c}}}})-1$ and $\gamma^{\delta_{\text{c}}}_{k}=(1/\varepsilon_{k}^{\text{MMSE},{\delta_{\text{c}}}})-1$. The common and private rates at user-$k$ are thus $R^{\delta_{\text{c}}}_{\text{c},k}=\log_2(1+\gamma^{\delta_{\text{c}}}_{\text{c},k})=-\log_2(\varepsilon_{\text{c},k}^{\text{MMSE},{\delta_{\text{c}}}})$ and $R^{\delta_{\text{c}}}_{k}=\log_2(1+\gamma^{\delta_{\text{c}}}_{k})=-\log_2(\varepsilon_{k}^{\text{MMSE},{\delta_{\text{c}}}})$.\par 
By allocating positive weights $(w^{\delta_{\text{c}}}_{\text{c},k},w^{\delta_{\text{c}}}_k)$ to user-$k$'s common and private rates, we define the augmented WMSEs as $\xi^{\delta_{\text{c}}}_{\text{c},k}\triangleq w^{\delta_{\text{c}}}_{\text{c},k}\varepsilon^{\delta_{\text{c}}}_{\text{c},k}-\log_2(w^{\delta_{\text{c}}}_{\text{c},k}),  \xi^{\delta_{\text{c}}}_{k}\triangleq w^{\delta_{\text{c}}}_{k}\varepsilon^{\delta_{\text{c}}}_{k}-\log_2(w^{\delta_{\text{c}}}_{k})$.
After optimizing over the equalizers and weights, the rate-WMMSE relationships are established as
\begin{equation}
\begin{split}
\xi_{\text{c},k}^{\text{MMSE},{\delta_{\text{c}}}}&\triangleq\min\limits_{w^{\delta_{\text{c}}}_{\text{c},k},g^{\delta_{\text{c}}}_{\text{c},k}}\xi^{\delta_{\text{c}}}_{\text{c},k}=1-R^{\delta_{\text{c}}}_{\text{c},k}\\
\xi_{k}^{\text{MMSE},{\delta_{\text{c}}}}&\triangleq\min\limits_{w^{\delta_{\text{c}}}_{k},g^{\delta_{\text{c}}}_{k}}\xi^{\delta_{\text{c}}}_{k}=1-R^{\delta_{\text{c}}}_{k},
\end{split}\label{ratemmserelation}
\end{equation}
where the optimal equalizers are 
$g_{\text{c},k}^{{\delta_{\text{c}}}*}=g_{\text{c},k}^{\text{MMSE},{\delta_{\text{c}}}}$ and $g_k^{{\delta_{\text{c}}}*}=g_k^{\text{MMSE},{\delta_{\text{c}}}}$, and the optimal weights are $w_{\text{c},k}^{{\delta_{\text{c}}}*}=w_{\text{c},k}^{\text{MMSE},{\delta_{\text{c}}}}=\left(\varepsilon_{\text{c},k}^{\text{MMSE},{\delta_{\text{c}}}}\right)^{-1}$ and $w_{k}^{{\delta_{\text{c}}}*}=w_{k}^{\text{MMSE},{\delta_{\text{c}}}}=\left(\varepsilon_{k}^{\text{MMSE},{\delta_{\text{c}}}}\right)^{-1}$. They are obtained by meeting the first-order optimality conditions.\par 
Using the rate-WMMSE relationships \eqref{ratemmserelation}, we then reformulate \eqref{updatevorigin} as 
\begin{equation}\label{WMMSE}
\begin{split}
\min\limits_{\begin{array}{l}{\bf c},{\bf p}_{\text{vec}}\\{\bf w}^{\delta_{\text{c}}},{\bf g}^{\delta_{\text{c}}}\end{array}} &-(1-\lambda)\sum_{k\in \mathcal{K}}\mu_k\left(-C_k+\xi^{\delta_{\text{c}}}_k({\bf p}_{\text{vec}})\right)\\
&+\frac{\rho}{2}\lVert{\bf p}_{\text{vec}}-{\bf D}_{\text{p}}{\bf u}^t+{\bf d}^t\lVert_2^2\\
s.t. \quad& \sum_{k'\in \mathcal{K}}C_{k'}+\xi^{\delta_{\text{c}}}_{\text{c},k}({\bf p}_{\text{vec}})\leq 1, \forall k\in\mathcal{K}\\
&{\bf p}_{\text{vec}}^H{\bf p}_{\text{vec}}\leq P_{\text{t}},\\
&{\bf c}\geq 0,
\end{split}
\end{equation}
where ${\bf w}^{\delta_{\text{c}}}=\left[w^{\delta_{\text{c}}}_1,\dots,w^{\delta_{\text{c}}}_K,w^{\delta_{\text{c}}}_{\text{c},1},\dots,w^{\delta_{\text{c}}}_{\text{c},K}\right]$ is the vector of all MSE weights. ${\bf g}^{\delta_{\text{c}}}=\left[g^{\delta_{\text{c}}}_1,\dots,g^{\delta_{\text{c}}}_K,g^{\delta_{\text{c}}}_{\text{c},1},\dots,g^{\delta_{\text{c}}}_{\text{c},K}\right]$ is the vector of all equalizers.\par 
Note that when ${\bf w}^{\delta_{\text{c}}}$ and ${\bf g}^{\delta_{\text{c}}}$ are fixed, \eqref{WMMSE} is a convex Quadratic Constrained Quadratic Programming (QCQP) according to $\{{\bf p}_{\text{vec}},{\bf c}\}$, which can be efficiently solved by CVX toolbox\cite{cvx}. The WMMSE algorithm to solve \eqref{updatev} is summarized in Algorithm 2. $\epsilon_0$ is the tolerance of this algorithm. The weights, equalizers, and precoders are iteratively updated till convergence. Readers are referred to \cite{mao2019RSWIPT} for more details on the WMMSE algorithm.
\begin{algorithm}\label{aowmmse}
	\caption{WMMSE algorithm}
	\LinesNumbered 
	\KwIn{$ {\bf u}^t, {\bf d}^t$.}
	\KwOut{${\bf v}^{t+1}$.}
	$k\leftarrow0$;\\
	${\bf p}_{\text{vec}}^{[k]}={\bf D}_{\text{p}}{\bf u}^t,{\bf c}^{[k]}=({\bf I}_{(K+2)N_{\text{t}}}-{\bf D}_{\text{p}}){\bf u}^{t}$;\\ $\text{WSR}^{[k]}$ is calculated from ${\bf p}_{\text{vec}}^{[k]}$ and ${\bf c}^{[k]}$;\\
	\Repeat{$\vert\text{WSR}^{[k]}-\text{WSR}^{[k-1]}\vert\leq\epsilon_1$}{
	${\bf w}^{\delta_{\text{c}}*}\leftarrow{\bf w}^{\text{MMSE},\delta_{\text{c}}}({\bf p}_{\text{vec}}^{[k]})$;\\
	${\bf g}^{\delta_{\text{c}}*}\leftarrow{\bf g}^{\text{MMSE},\delta_{\text{c}}}({\bf p}_{\text{vec}}^{[k]})$;\\
	update $({\bf p}_{\text{vec}}^{[k+1]},{\bf c}^{[k+1]})$ by solving QCQP \eqref{WMMSE} with the updated ${\bf w}^{\delta_{\text{c}}*},{\bf g}^{\delta_{\text{c}}*}$;\\
	update $\text{WSR}^{[k+1]}$ by using ${\bf p}_{\text{vec}}^{[k+1]}$ and ${\bf c}^{[k+1]}$\\
	$k++$;
	}
	${\bf v}^{t+1}=[{\bf c}^{[k]}; {\bf p}_{\text{vec}}^{[k]}]$;
\end{algorithm}
\subsection{MM-based Algorithm for ${\bf u}$-update}
In this subsection, we propose an approach based on MM algorithm to solve \eqref{updateu}. 
By defining ${\bf p}_{\text{u}}={\bf D}_{\text{p}}{\bf u}$ and 
\begin{equation}
\begin{split}
{\bf D}_{\text{p},k}=\left[{\bf 0}^{N_{\text{t}}\times(k-1)N_{\text{t}}}\quad{\bf I}_{N_{\text{t}}}\quad {\bf 0}^{N_{\text{t}}\times(K+2-k)N_{\text{t}}}\right],\\ 
k=1,2,\dots,K+2,
\end{split}\label{DPK}
\end{equation} \eqref{updateu} is equivalently formulated as
\begin{equation}
\begin{split}\label{updateuorigin}
\min_{{\bf p}_{\text{u}}}\quad&\lambda\sum_{m=1}^{M}\lvert P_{\text{d}}(\theta_m)-{\bf a}^H(\theta_m)\big(\sum_{k=1}^{K+2}{\bf D}_{\text{p},k}{\bf p}_{\text{u}}{{\bf p}_{\text{u}}}^H{\bf D}_{\text{p},k}^H\big) \\
&{\bf a}(\theta_m)\lvert ^2+\frac{\rho}{2}\lVert{\bf D}_{\text{p}}{\bf v}^{t+1}-{\bf p}_{\text{u}}+{\bf d}^t\rVert_2^2\\
s.t. \quad &\text{diag}(\sum_{k=1}^{K+2}{\bf D}_{\text{p},k}{\bf p}_{\text{u}}{{\bf p}_{\text{u}}}^H{\bf D}_{\text{p},k}^H)=\frac{P_{\text{t}}{\bf 1}^{N_{\text{t}}\times 1}}{N_{\text{t}}}.
\end{split}
\end{equation} 
Since the complicated quartic formulation makes it difficult to find a tight relaxation of \eqref{updateu}, we propose an MM-based iterative approach to solve this problem referring to \cite{MMTSP}.\par 
Recall Lemma 1 in \cite{Song2015Optimization} that will be used in our MM-based algorithm.
\newtheorem{lemma}{Lemma}
\begin{lemma}\label{lemma1}
	Let ${\bf L}$, ${\bf M}$ be the $n\times n$ Hermitian matrices and ${\bf M}\succeq {\bf L}$. For any point ${\bf x}_0\in\mathbb{C}^n$, there is ${\bf x}^H{\bf L}{\bf x}\leq {\bf x}^H{\bf M}{\bf x}+2\mathfrak{R}\{{\bf x}^H({\bf L-M}){\bf x}_0\}+{\bf x}_0^H({\bf M-L}){\bf x}_0$.$\hfill\blacksquare$
\end{lemma}\par 
Minimizing the objective function of \eqref{updateuorigin} is equivalent to minimizing
\begin{equation}\label{objorigin}
f_{\text{u}}({\bf p}_{\text{u}})=\lambda\sum_{m=1}^{M}\lvert a_m-{\bf p}_{\text{u}}^H{\bf Z}_m{\bf p}_{\text{u}}\lvert^2-\rho\mathfrak{R}\{{\bf p}_{\text{u}}^H\hat{\bf d}^t\},
\end{equation} 
where $a_m=P_{\text{d}}(\theta_m)$, $\hat{\bf d}^t={\bf D}_{\text{p}}{\bf v}^{t+1}+{\bf d}^t$, ${\bf Z}_m=\sum_{k=1}^{K+2}{\bf D}_{\text{p},k}^H{\bf a}(\theta_m){\bf a}^H(\theta_m){\bf D}_{\text{p},k}$.
Let $q_m({\bf p}_{\text{u}})={\bf p}_{\text{u}}^H{\bf Z}_m{\bf p}_{\text{u}}$ and ${\bf Q}_m=\text{vec}({\bf Z}_m)\text{vec}({\bf Z}_m)^H$,
we use Lemma \ref{lemma1} and get
\begin{equation}
\begin{split}
&q_m({\bf p}_{\text{u}})^2\leq 2\lambda_{\text{max}}({\bf Q}_m)P_{\text{t}}^2-q_m({\bf p}_{\text{u}}^k)^2\\
&+2\mathfrak{R}\{q_m({\bf p}_{\text{u}}^k)q_m({\bf p}_{\text{u}})-\lambda_{\text{max}}({\bf Q}_m)({\bf p}_{\text{u}}^H{\bf p}_{\text{u}}^k({\bf p}_{\text{u}}^k)^H{\bf p}_{\text{u}})\}.
\end{split}\label{1mmfinal}
\end{equation}
Substituting \eqref{1mmfinal} into \eqref{objorigin}, we then have the majorization function of \eqref{objorigin} as
\begin{equation}
\begin{split}
f_{\text{u}}({\bf p}_{\text{u}})\leq f_{\text{u}}^1({\bf p}_{\text{u}},{\bf p}_{\text{u}}^k)=C_{\text{u}}+{\bf p}_{\text{u}}^H{\bf Q'}{\bf p}_{\text{u}}-\rho\mathfrak{R}\{{\bf p}_{\text{u}}^H\hat{\bf d}^t\}
\end{split}\label{1mmfinish}
\end{equation}
where 
$
C_{\text{u}}=\lambda\sum_{m=1}^{M}\big[2\lambda_{\text{max}}({\bf Q}_m)P_{\text{t}}^2-q_m({\bf p}_{\text{u}}^k)^2+a_m^2\big]$, $
{\bf Q'}=\lambda\sum_{m=1}^{M}\big[2\big(q_m({\bf p}_{\text{u}}^k)-a_m\big){\bf Z}_m-2\lambda_{\text{max}}({\bf Q}_m){\bf p}_{\text{u}}^k({\bf p}_{\text{u}}^k)^H\big]$.

It is obvious that $C_{\text{u}}$ and ${\bf Q'}$ are constant when ${\bf p}_{\text{u}}^k$ is determined in each iteration. Since that there is still a quadratic term in \eqref{1mmfinish}, we apply Lemma \ref{lemma1} again and have
\begin{equation}
\begin{split}
{\bf p}_{\text{u}}^H{\bf Q'}{\bf p}_{\text{u}}
\leq&\lambda_{\text{max}}({\bf Q}')P_{\text{t}}+({\bf p}_{\text{u}}^k)^H(\lambda_{\text{max}}({\bf Q}'){\bf I}-{\bf Q'}){\bf p}_{\text{u}}^k\\
&+2\mathfrak{R}\{{\bf p}_{\text{u}}^H({\bf Q'}-\lambda_{\text{max}}({\bf Q}'){\bf I}){\bf p}_{\text{u}}^k\}.
\end{split}
\end{equation}
Therefore, we obtain
\begin{equation}
\begin{split}
&f_{\text{u}}({\bf p}_{\text{u}})\leq f_{\text{u}}^1({\bf p}_{\text{u}},{\bf p}_{\text{u}}^k)\leq f_{\text{u}}^2({\bf p}_{\text{u}},{\bf p}_{\text{u}}^k)\\
&=C_{\text{u}}'+\mathfrak{R}\big\{{\bf p}_{\text{u}}^H\big[2({\bf Q'}-\lambda_{\text{max}}({\bf Q}'){\bf I}){\bf p}_{\text{u}}^k-\rho\hat{\bf d}^t\big]\big\}
\end{split}
\end{equation}
where 
\begin{equation}
C_{\text{u}}'=C_{\text{u}}+\lambda_{\text{max}}({\bf Q}')P_{\text{t}}+({\bf p}_{\text{u}}^k)^H(\lambda_{\text{max}}({\bf Q}'){\bf I}-{\bf Q'}){\bf p}_{\text{u}}^k.
\end{equation}\par 
We now obtain the linear majorization function of the quartic objective function $f_{\text{u}}({\bf p}_{\text{u}})$ at the point ${\bf p}_{\text{u}}^k$ as $f_{\text{u}}^2({\bf p}_{\text{u}},{\bf p}_{\text{u}}^k)$. After omitting the constant term, the majorized problem of \eqref{updateuorigin} is denoted as
\begin{equation}
\begin{split}\label{aftermm}
\min_{{\bf p}_{\text{u}}}\quad&\mathfrak{R}\big\{{\bf p}_{\text{u}}^H\big[2({\bf Q'}-\lambda_{\text{max}}({\bf Q}'){\bf I}){\bf p}_{\text{u}}^k-\rho\hat{\bf d}^t\big]\big\}\\
s.t. \quad &\text{diag}(\sum_{k=1}^{K+1}{\bf D}_{\text{p},k}{\bf p}_{\text{u}}{{\bf p}_{\text{u}}}^H{\bf D}_{\text{p},k}^H)=\frac{P_{\text{t}}{\bf 1}^{N_{\text{t}}\times 1}}{N_{\text{t}}}.
\end{split}
\end{equation} 
To further simplify \eqref{aftermm}, we denote 
$
\hat{\bf k}'={\bf f}_{\hat{\text{k}}}({\bf p}_{\text{u}}^k)=-2({\bf Q'}-\lambda_{\text{max}}({\bf Q}'){\bf I}){\bf p}_{\text{u}}^k-\rho\hat{\bf d}^t,\label{k'}
$ and split ${\bf p}_{\text{u}}$ into $N_{\text{t}}$ sub-vectors as  
\begin{equation}
\begin{split}
\hat{\bf p}_j=\left[[p_{\text{u}}]_j,[p_{\text{u}}]_{N_{\text{t}}+j},\dots, [p_{\text{u}}]_{((K+1)N_{\text{t}}+j}\right]^T,\\ j=1, 2,\dots, N_{\text{t}},
\end{split}\label{pjfromp}
\end{equation}
where $[p_{\text{u}}]_j$ is the $j$th entry in ${\bf p}_{\text{u}}$. We then split $\hat{\bf k}'$ by the same means into
$
\hat{\bf k}_j=\left[\hat{k}'_j,\hat{k}'_{N_{\text{t}}+j},\dots, \hat{k}'_{(K+1)N_{\text{t}}+j}\right]^T, j=1, 2,\dots, N_{\text{t}}.	
$
where $\hat{k}'_j$ is the $j$th entry in $\hat{\bf k}'$. 
According to \cite{xU2020Access}, \eqref{aftermm} has a closed-form solution as 
\begin{equation}
\hat{\bf p}_{j}^*=\frac{\sqrt{\frac{P_{\text{t}}}{N_{\text{t}}}}}{\lVert\hat{\bf k}_{j}\lVert_2}\hat{\bf k}_{j}, j=1, 2,\dots, N_{\text{t}},\label{finalMMsol}
\end{equation}
and ${\bf p}_{\text{u}}^*$ can be recovered from $\hat{\bf p}_{j}^*$ according to \eqref{pjfromp}. \par 
Therefore, \eqref{updateuorigin} is solved by iteratively computing $\hat{\bf p}_{j}^*$ based on \eqref{finalMMsol} until convergence. The initial point for the proposed MM-based algorithm can be chosen via classic Semi-definite Relaxation (SDR) \cite{Luo2010Semidefinite}.
The MM-based algorithm is summarized in Algorithm 3, where $\epsilon_2$ is its tolerance. The first $K$ elements of the output ${\bf u}^{t+1}$ are directly copied from ${\bf v}^{t+1}$ as they do not participate in the optimization \eqref{updateuorigin} and have no relations with the whole ${\bf u}$-update. \par 
\begin{algorithm}
	\caption{MM-based algorithm}
	\LinesNumbered 
	\KwIn{${\bf v}^{t+1}, {\bf d}^{t}$, $k\leftarrow0$, ${\bf p}_{\text{u}}^{[k]}$.}
	\KwOut{${\bf u}^{t+1}$.}
	\Repeat{$\lVert{\bf p}_{\text{u}}^{[k]}-{\bf p}_{\text{u}}^{[k-1]}\lVert_2\leq\epsilon_2$}{
	$\hat{\bf k}'={\bf f}_{\hat{\text{k}}}({\bf p}_{\text{u}}^{[k]})$;\\
	\For{$j=1$ to $N_{\text{t}}$}{$\hat{\bf k}_j=\left[\hat{k}'_j,\hat{k}'_{N_{\text{t}}+j},\dots, \hat{k}'_{(K+1)N_{\text{t}}+j}\right]^T$;\\
	$\hat{\bf p}_{j}^*=\frac{\sqrt{\frac{P_{\text{t}}}{N_{\text{t}}}}}{\lVert\hat{\bf k}_{j}\lVert_2}\hat{\bf k}_{j}$;}
	${\bf p}_{\text{u}}^{[k+1]}$ is obtained by doing the inverse operation of \eqref{pjfromp} using $\hat{\bf p}_{j}^*$;\\
	$k++$;
	}
	${\bf u}^{t+1}=[({\bf I}_{(K+2)N_{\text{t}}}-{\bf D}_{\text{p}}){\bf v}^{t+1};{\bf p}_{\text{u}}^{[k]}]$;
\end{algorithm}

\subsection{Computational Complexity Analysis}
In this part, we analyse the complexity of the proposed ADMM-based framework to solving \eqref{CR origin}. From Algorithm 1, it is clear that the complexity in one iteration consists of running the WMMSE algorithm and the MM-based algorithm.  \par 
According to \cite{nesterov1994interior}, the polynomial complexity of implementing WMMSE-AO to solve the QCQP problem \eqref{WMMSE} via interior-point method is $\mathcal{O}([3+(K+2)(N_{\text{t}})]\times [(K+2)(N_{\text{t}})]^2)=\mathcal{O}([(K+2)(N_{\text{t}})]^3)$. \par 
For MM-based algorithm, 
the complexity of calculating the largest eigenvalue of ${\bf Q}'$ in each MM iteration is $\mathcal{O}([(K+2)(N_{\text{t}})]^{3})$, while that for other steps of linear operation is $\mathcal{O}([(K+2)(N_{\text{t}})]^{2})$. Hence, the complexity of the MM-based iteration is $\mathcal{O}([(K+2)(N_{\text{t}})]^{3})$. \par 
Therefore, for the proposed DFRC, the complexity of each iteration of our ADMM-based framework is polynomial as $\mathcal{O}([(K+2)(N_{\text{t}})]^{3})$. Following the same analysis, we can obtain the complexity of RSMA-assisted DFRC without radar sequence as $\mathcal{O}([(K+1)(N_{\text{t}})]^{3})$.\par

\section{Numerical Results}
In this section, we provide numerical results to validate the performance of the proposed RSMA-assisted DFRC. We mainly reveal the superior advantages of using RSMA over SDMA in DFRC and further demonstrate that the common stream of RSMA outperforms radar sequence. TDRC and FDRC are also compared as baselines.
\subsection{Settings of Numerical Experiments}
 We assume the proposed DFRC employs a ULA where $N_{\text{t}}=8$ with half-wavelength spacing, i.e., $\Delta=0.5$. The total transmit power budget is $P_{\text{t}}=20\text{dBm}$ and the noise power at each user is $0\text{dBm}$. We consider four users and three targets of interest. These users are numbered as User-1 to User-4. We generate four channel vectors randomly obeying the i.i.d. standard complex Gaussian distribution to represent the four user channels. The three targets of interest are Target-1 with the azimuth angle range $[-6,6]$\textdegree, Target-2 at $[-56,-44]$\textdegree and Target-3 at $[44,56]$\textdegree. All four users and their channel vectors, together with the three targets will be chosen and combined by different means in the following experiments for thorough analysis. TDRC and FDRC are included as baselines and we assume the radar and communication function share the same power budget $P_{\text{t}}/2$ in FDRC.\par 
For the optimization variables, precoders $[{\bf{p}}_{\text{c}},{\bf p}_1,\dots,{\bf{p}}_K]$ are initialized through Maximum Ratio Combining (MRC) method following \cite{joudeh2016sum}, and ${\bf p}_{\text{r}}$ and ${\bf d}$ are initialized randomly. ${\bf c}$ is initialized as ${\bf 1}^{K\times 1}$. The desired beampattern is generated following the matching method in \cite{stoica2007probing}. \par 
In the following results, in order to achieve a clearer comparison, we use Root Mean Squared Error (RMSE) instead of MSE for beampattern approximation. In the figures, ``RSMA, no Rad-Seq, DFRC" represents the scheme ``RSMA-assisted DFRC without radar sequence", ``RSMA, noSIC Rad-Seq, DFRC" represents ``RSMA-assisted DFRC with radar sequence", and ``RSMA, SIC Rad-Seq, DFRC" represents ``RSMA-assisted DFRC with radar sequence and SIC". The three SDMA-assisted DFRC modes are denoted in the same manner.
\subsection{Metric for Radar-to-Communication Interference: LB-IBR} 
Since we adopt WSR and beampattern approximation MSE as optimization objectives, they are unable to quantitatively depict the dual-functional interference. \par From this perspective, we define Interference-to-Beampattern-Ratio (IBR) and derive its lower bound as LB-IBR. The definitions are based on SDMA-assisted DFRC without radar sequence, such that the advantage of enabling the common stream and radar sequence can be analysed accordingly. IBR is defined as follows.
\begin{equation}
\text{IBR}(\theta_m)=\frac{\sum_{k\in \mathcal{K}}\sum_{j\neq k}{\bf h}_j^H{\bf p}_k{\bf p}_k^H{\bf h}_j}{\sum_{k\in \mathcal{K}}{\bf a}(\theta_m)^H{\bf p}_k{\bf p}_k^H{\bf a}(\theta_m)}.\label{IBR}
\end{equation}
The denominator is the transmit beampattern value of SDMA-assisted DFRC without radar sequence at $\theta_m$, while the numerator is the total interference leakage towards the communication users. A higher $\text{IBR}(\theta_m)$ means more interference will be leaked to users when a unit-power beampattern at $\theta_m$ is achieved, which reflects stronger radar-to-communication interference in DFRC.\par
Interestingly, by exploiting the precoder power constraint \eqref{CR origin c4}, we can obtain the lower bound of $\text{IBR}(\theta_m)$ that does not contain the optimization variable ${\bf p}_k$. This lower bound is denoted as LB-IBR and can be expressed as
\begin{equation}
\text{LB-IBR}(\theta_m)=(1/N_{\text{t}}^2)\min_{k\in \mathcal{K}}(\sum_{j\neq k}\lVert{\bf a}(\theta_m)^H{\bf h}_j\lVert_2^2)\le \text{IBR}(\theta_m).\label{LBIBR}
\end{equation}
%
%
The derivation is shown in Appendix. LB-IBR can be used to evaluate the basic radar-to-communication interference level for SDMA-assisted DFRC without radar sequence in the operating environment.\par 

\subsection{Beampattern Approximation Performance}
We first show the beampattern approximation performance of the proposed DFRC. We take the DFRC enabling RSMA and radar sequence with SIC as an example, where the DFRC serves all four users and detects all three targets.\par 
\begin{figure}[t]
	\centering
	\subfigure[$\lambda=1\times10^{-5}$, $\text{WSR}=3.85$bps/Hz, RMSE=5.5.] {\includegraphics[width=0.48\linewidth]{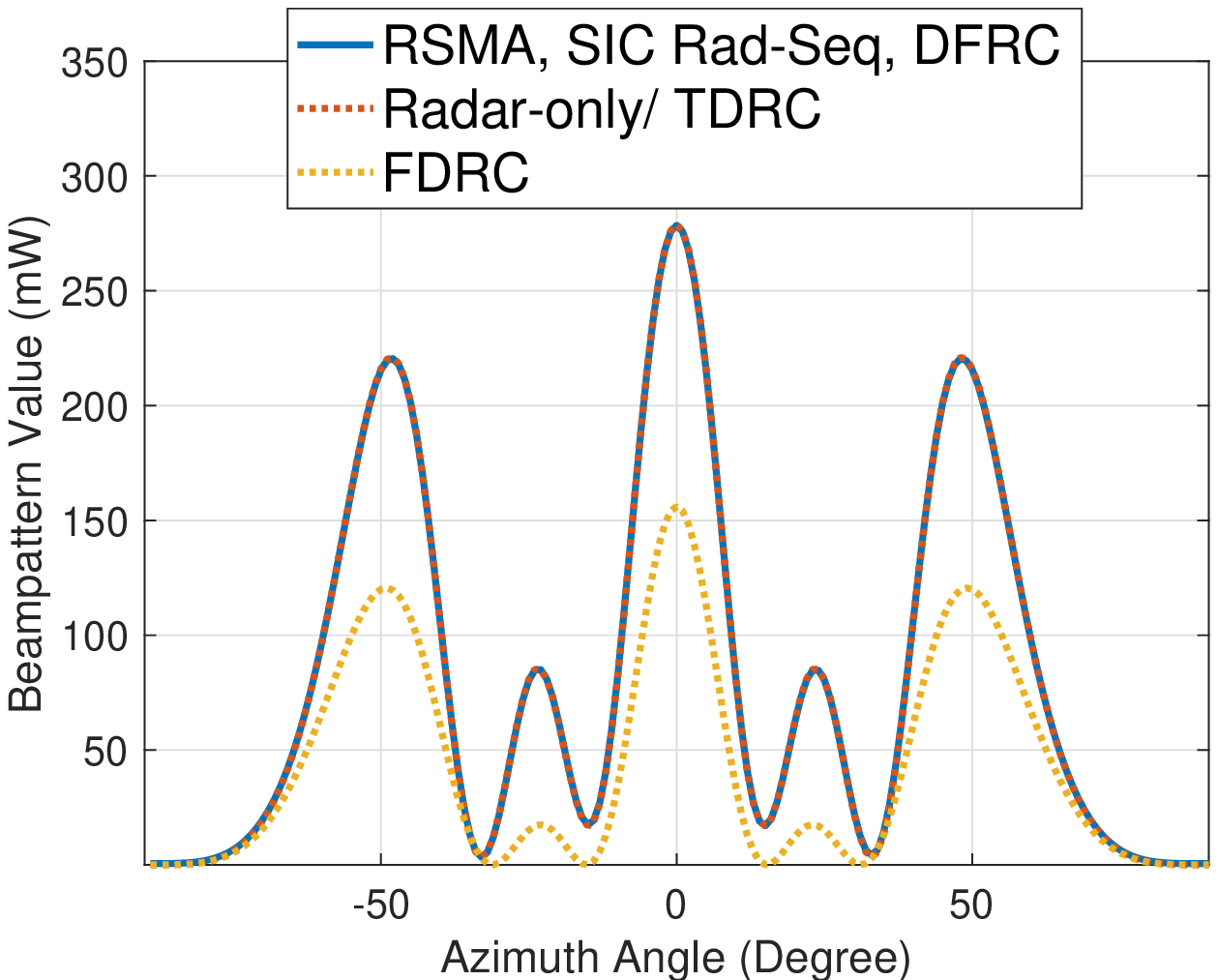}\label{beamfulllook1}}
	\subfigure[$\lambda=1\times10^{-6}$, $\text{WSR}=5.65$bps/Hz, RMSE=$4.7\times10^{2}$.] {\includegraphics[width=0.48\linewidth]{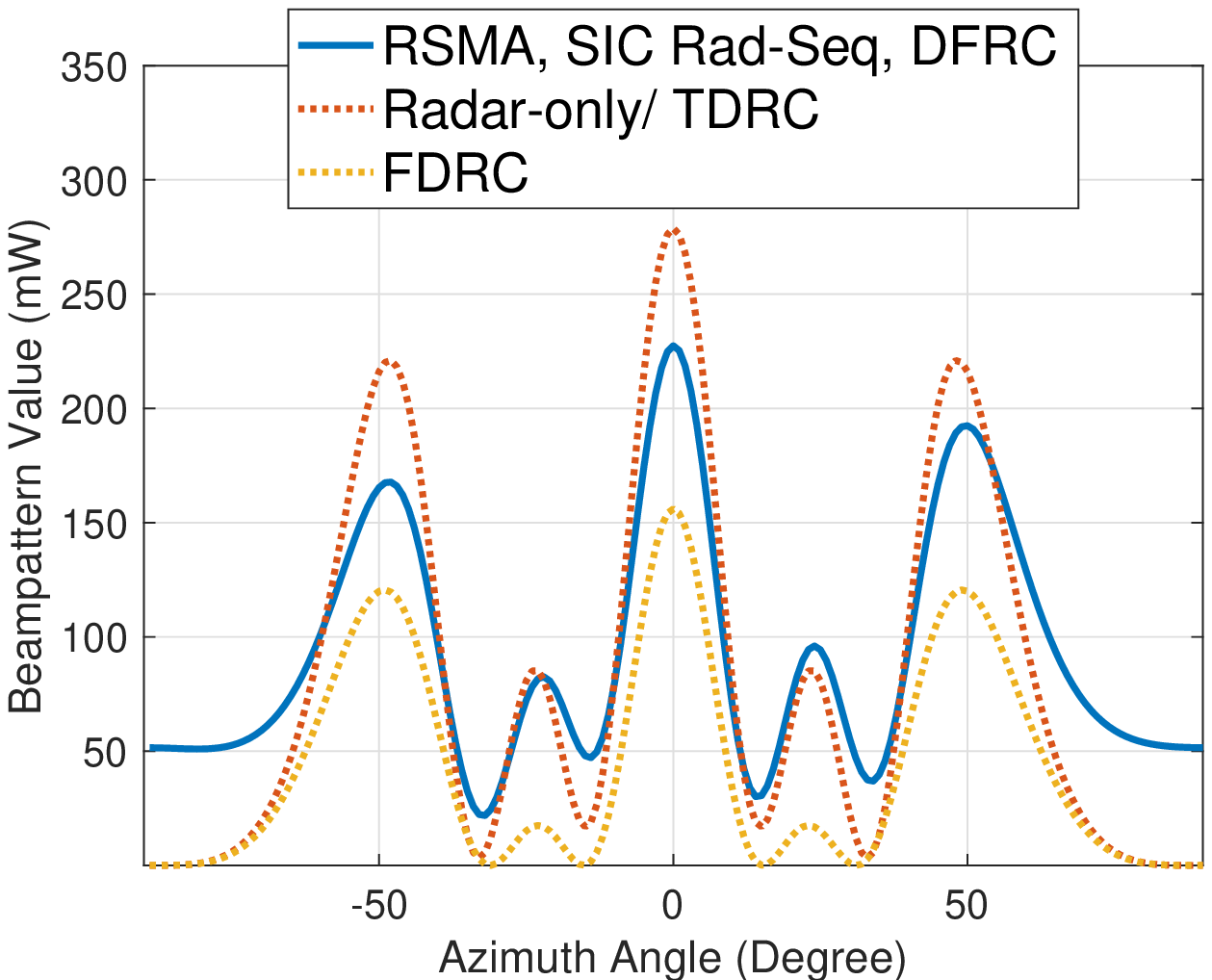}\label{beamfulllook2}}
	\caption{Beampattern approximation comparisons among proposed DFRC with different WSR and RMSE, MIMO radar, TDRC and FDRC.}\label{beamfulllook}
\end{figure}
Fig. \ref{beamfulllook} shows the transmit beampattern approximation performance of the proposed DFRC system compared with pure MIMO radar with $N_{\text{t}}$ antennas, TDRC and FDRC. The beampattern of DFRC is obtained by solving \eqref{CR origin}. The desired beampattern is obtained via beampattern matching methods for a pure MIMO radar, which can be achieved in TDRC. In Fig. \ref{beamfulllook1}, it is clear that DFRC is able to fully approximate MIMO radar's beampattern, and has near 3dB gain over FDRC. This is because the latter has only half the power budget when doing the radar function. Fig. \ref{beamfulllook2} shows that by decreasing the regularization parameter to $\lambda=1\times10^{-6}$, DFRC experiences near 1.8bps/Hz increase in WSR but also near 19.3dB increase in beampattern approximation RMSE. This reveals that there is a tradeoff between beampattern approximation MSE and WSR.\par  

\subsection{RSMA vs. SDMA in DFRC}
In this subsection, we analyse the advantages of enabling RSMA in DFRC compared with using SDMA in terms of beampattern approximation RMSE and WSR tradeoff performance. We also demonstrate how the common stream contributes to DFRC and explain why RSMA outperforms SDMA when user's channels, target location, user number and target number change.\par 
Too eliminate influence of radar sequence, we only investigate RSMA-assisted and SDMA-assisted DFRC without radar sequence in this part. Performance of both TDRC and FDRC is included as baselines in the simulation results. 
\subsubsection{RSMA Gain vs. Users' Channels}
To show how RSMA benefits DFRC when users' channels change, we first fix the user number, target number and the target location. We then set up the first scenario, denoted as Scenario-UC, where different user pairs are served, i.e., User-1 and User-4, User-2 and User-4, User-3 and User-4. In these three scenarios, only one target, i.e., Target-1, is supposed to be detected.\par 
\begin{figure}[t]
	\centering
	\subfigure[Tradeoffs when serving User-1 and User-4.] {\includegraphics[width=0.48\linewidth]{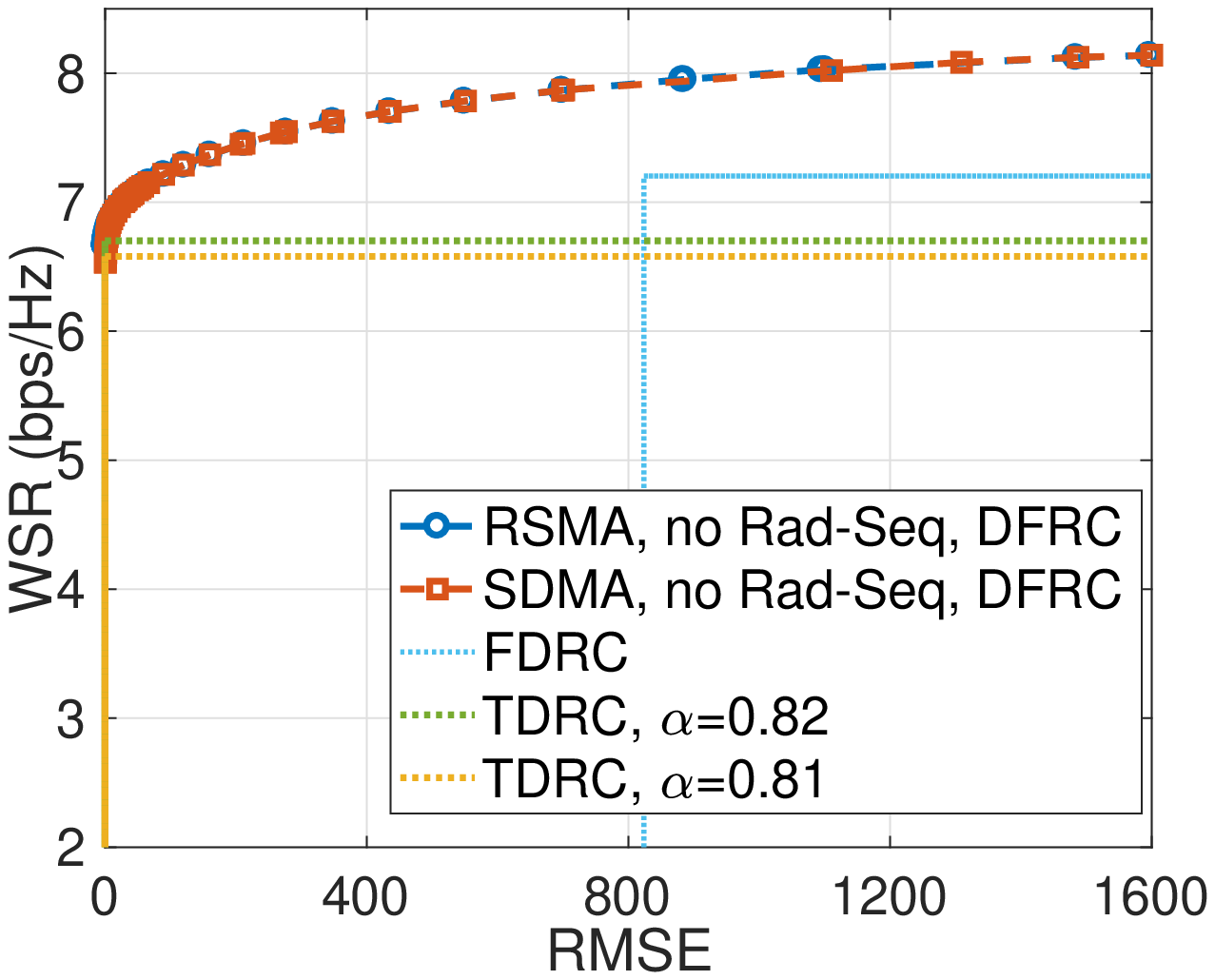}\label{5a}}
	\subfigure[Tradeoffs when serving User-2 and User-4.] {\includegraphics[width=0.48\linewidth]{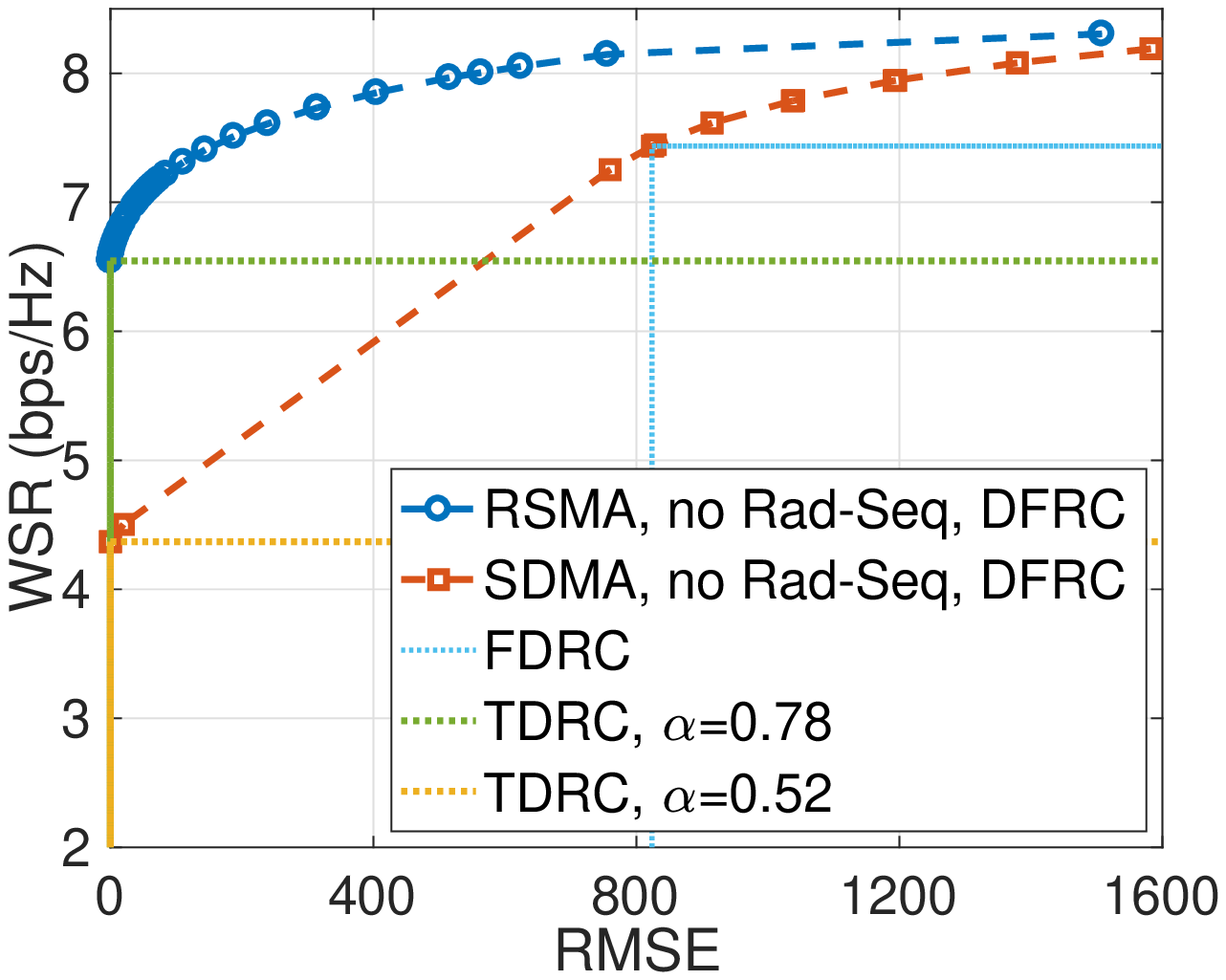}\label{5b}}
	\subfigure[Tradeoffs when serving User-3 and User-4.] {\includegraphics[width=0.48\linewidth]{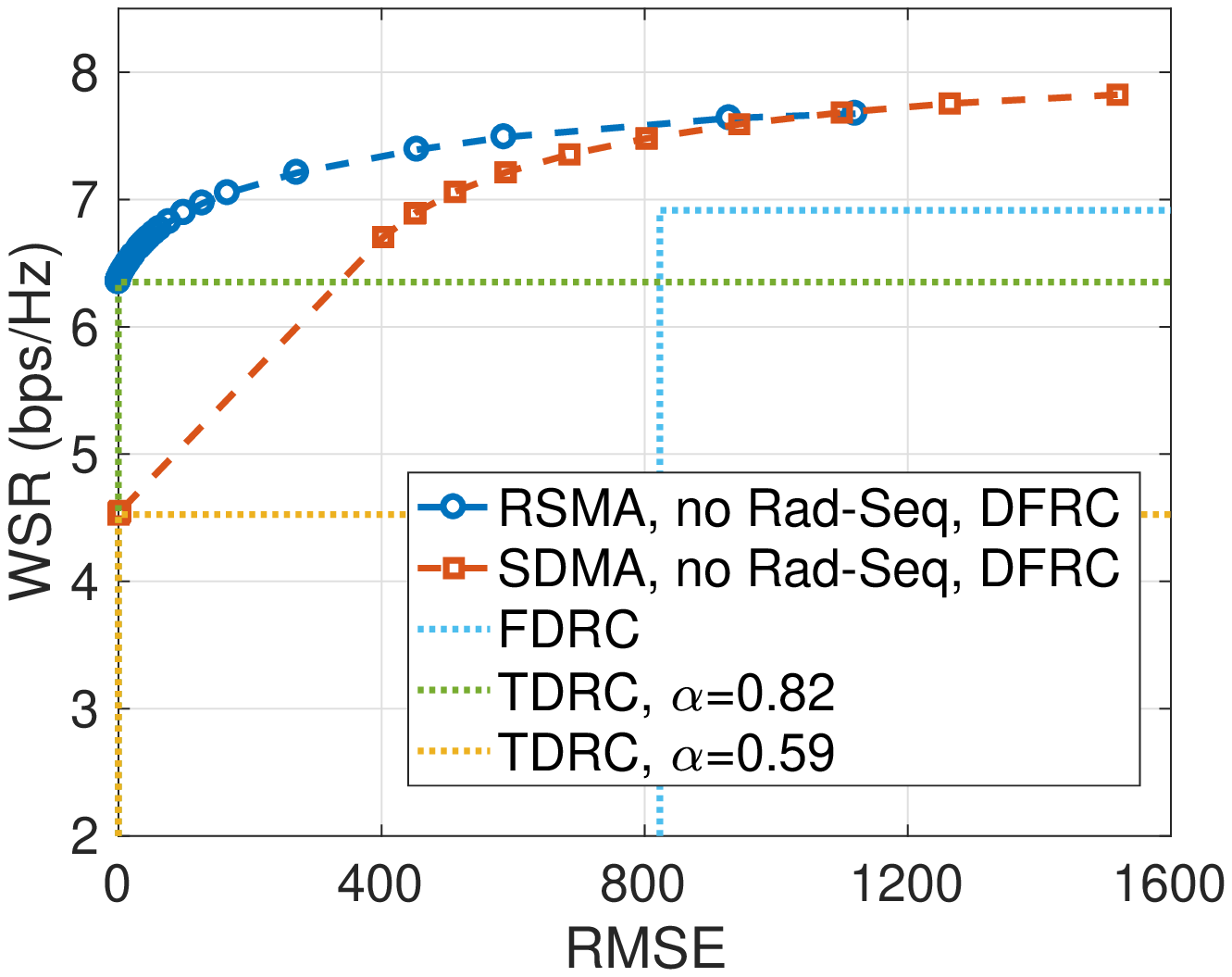}\label{5c}}
	\subfigure[LB-IBR when serving different user pairs.] {\includegraphics[width=0.48\linewidth]{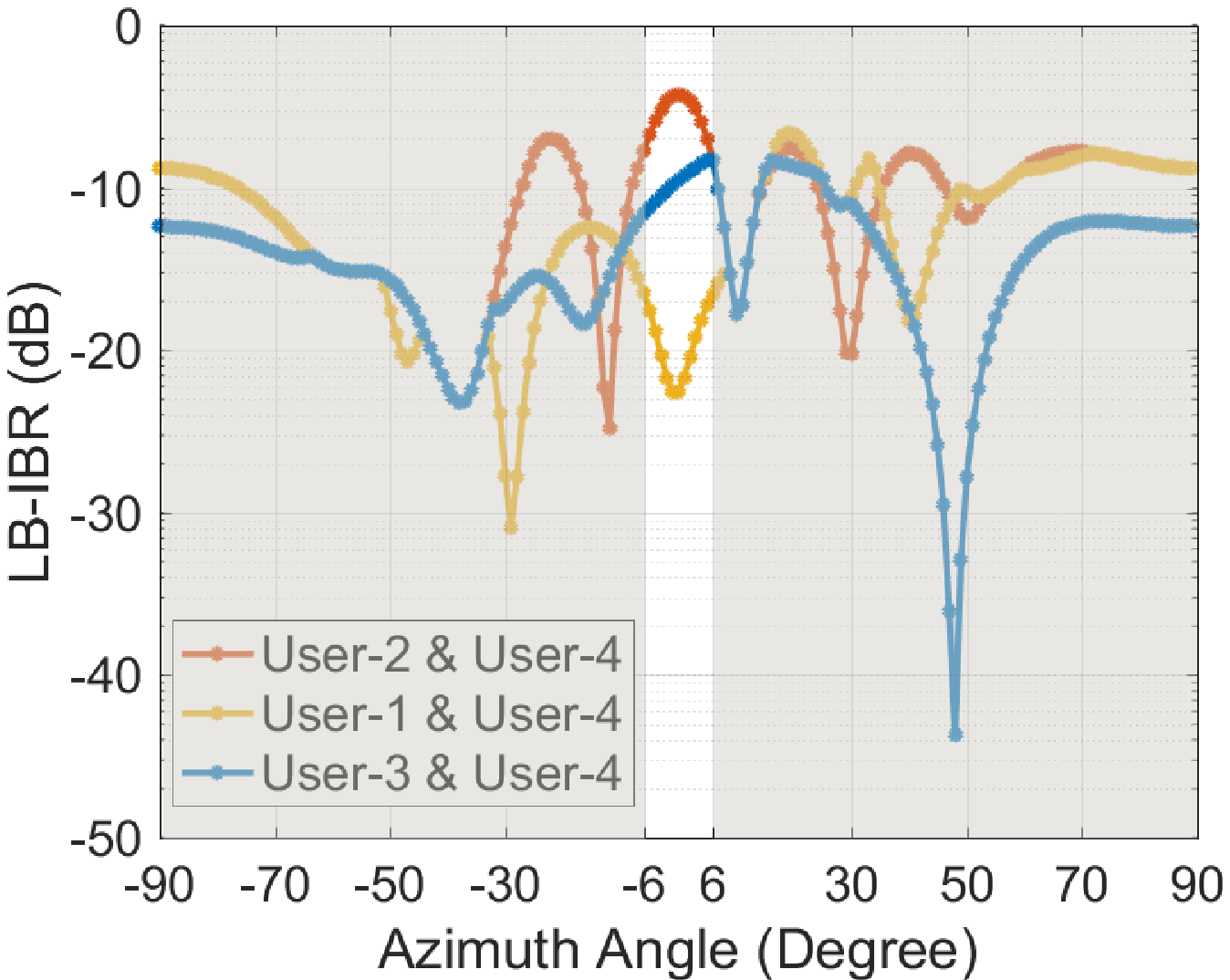}\label{UCIBR}}
	\caption{Tradeoff and LB-IBR for Scenario-UC.}\label{UC}
\end{figure}
In Fig. \ref{5a}--\ref{5c}, the tradeoffs between RMSE and WSE for RSMA-assisted and SDMA-assisted DFRC without radar sequence, FDRC and TDRC strategies are compared in this scenario. It is clear that RSMA-assisted DFRC without radar sequence outperforms FDRC with 0.8bps/Hz WSR gain on average given the same beampattern approximation RMSE in all scenarios, and surpasses TDRC when $\alpha$ is smaller than around 0.8. Additionally, RSMA-assisted DFRC without radar sequence outperforms SDMA-assisted DFRC without radar sequence as well, but the tradeoff gain varies according to different user pairs. Specifically, when serving User-2\&4, the former achieves the largest WSR gain of 1.2bps/Hz when RMSE$\le1$, while the WSR gain almost disappears when serving User-1\&4. Then we use LB-IBR to analyze the radar-to-communication interference level in this scenario, so as to explain why RSMA-assisted DFRC outperforms SDMA-assisted DFRC when radar sequence is disabled.\par
\begin{figure}[t]
	\centering
	\subfigure[Beampattern contribution when serving User-1 and User-4.] {\includegraphics[width=0.48\linewidth]{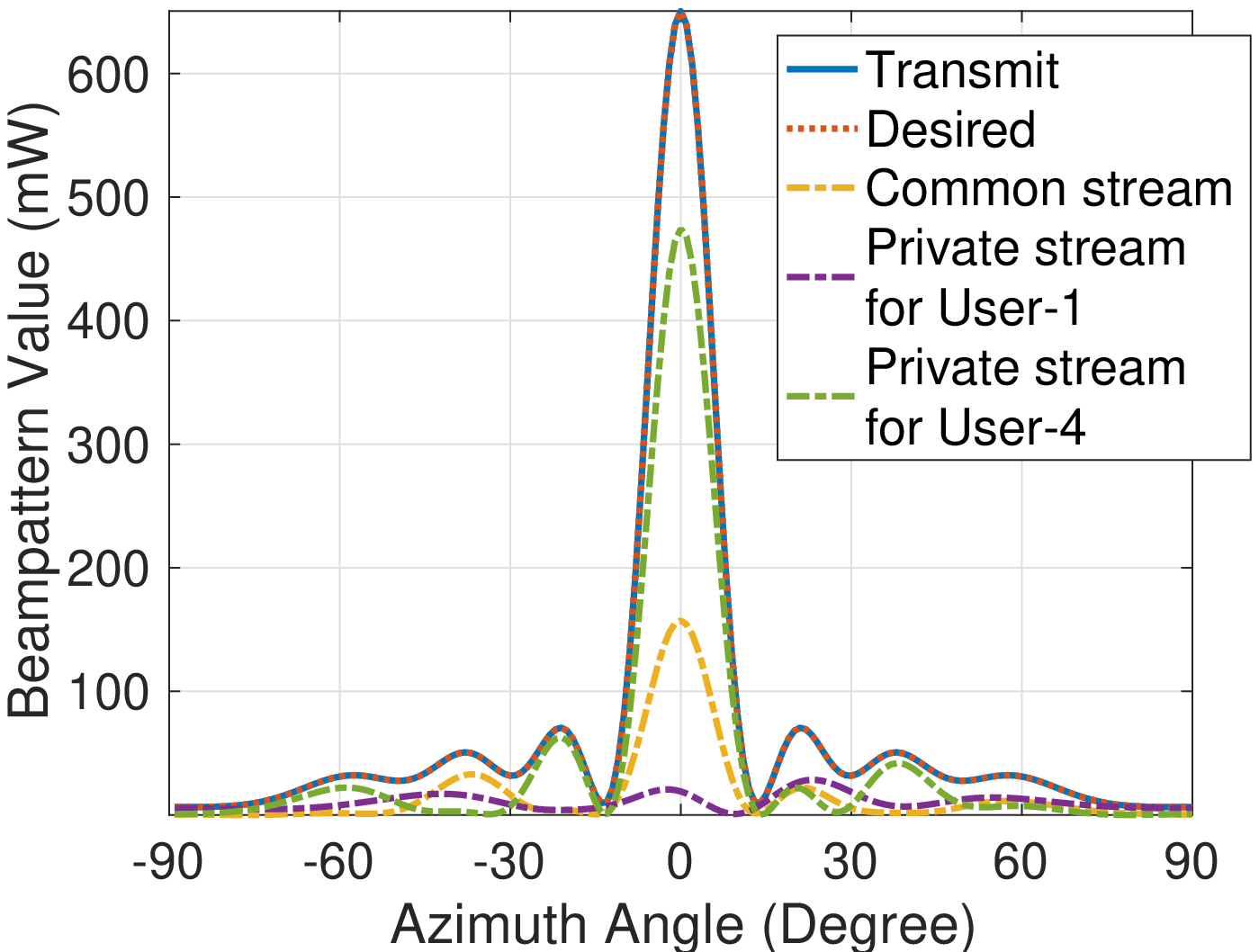}}
	\subfigure[Beampattern contribution when serving User-2 and User-4.] {\includegraphics[width=0.48\linewidth]{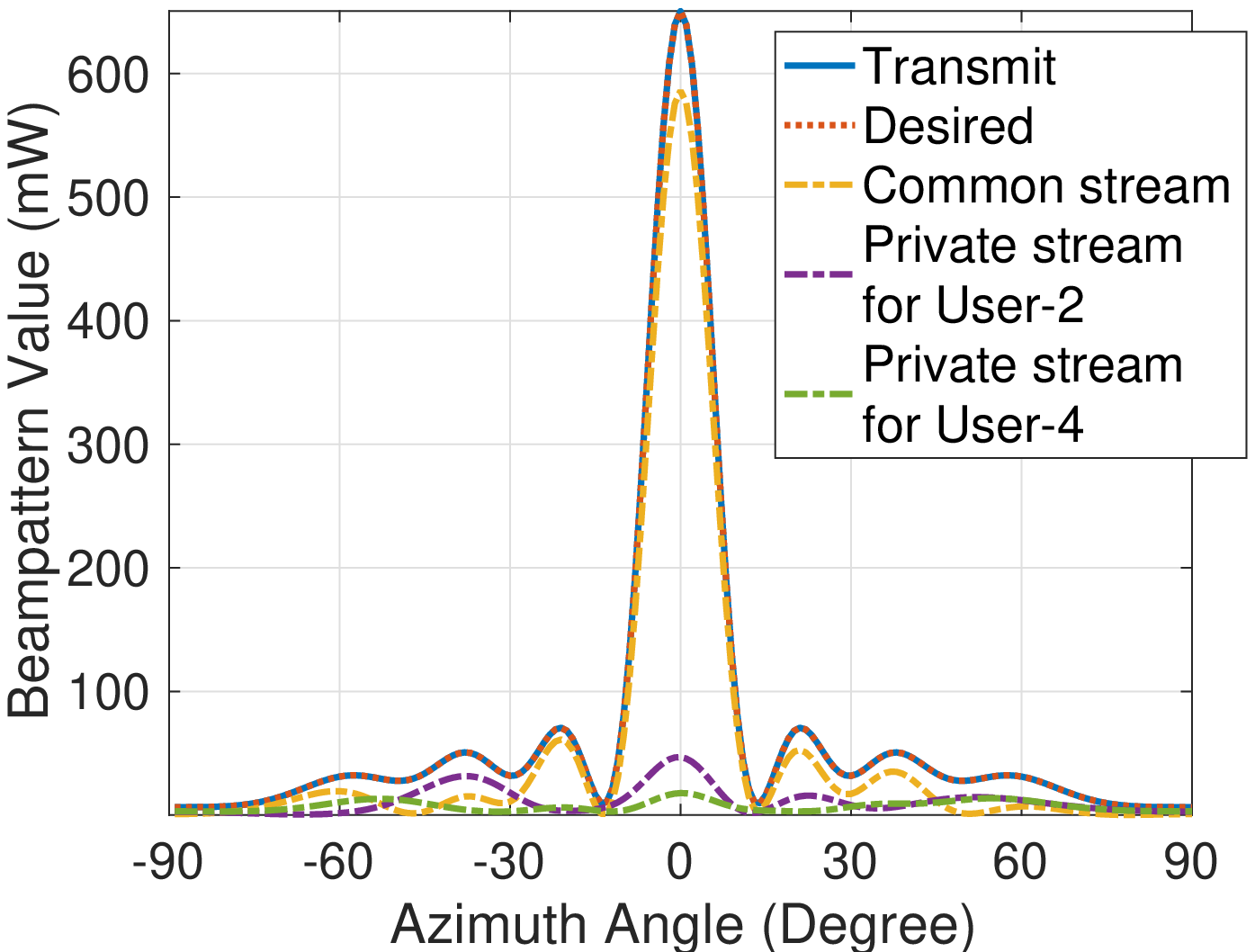}\label{4b}}
	\subfigure[Beampattern contribution when serving User-3 and User-4.] {\includegraphics[width=0.48\linewidth]{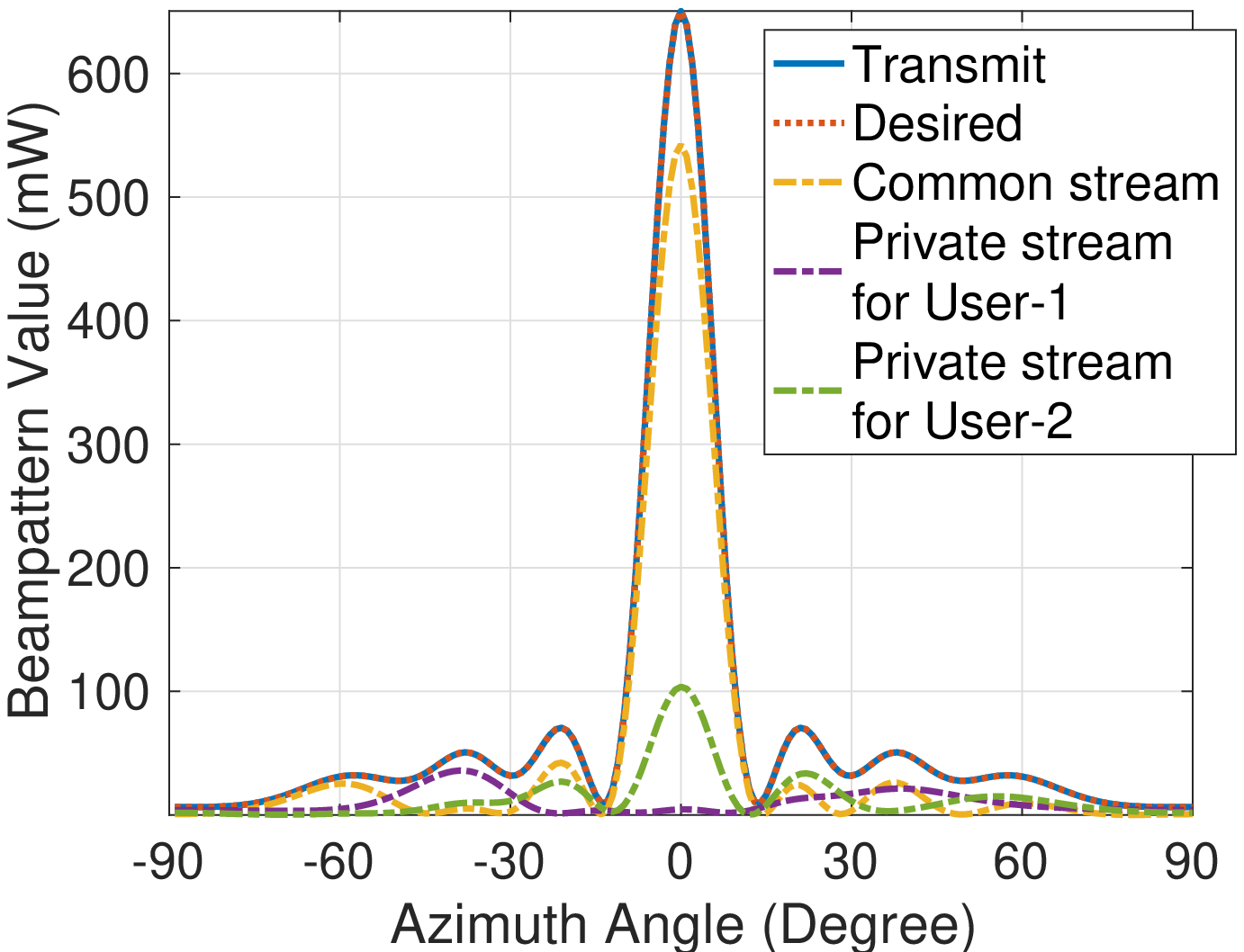}}
	\caption{Beampattern contribution of RSMA-assisted DFRC without radar sequence for Scenario-UC.}\label{BeamConUC}
\end{figure}

  
Fig. \ref{UCIBR} demonstrates the LB-IBR in Scenario-UC, where the azimuth angle range of Target-1 is highlighted. Obviously, at the location of Target-1, LB-IBR is the highest when User-2\&4 are served but lowest when User-1\&4 are served, which means detecting Target-1 leads to the highest and lowest interference on the former and latter user pair, respectively. Fig. \ref{BeamConUC} further displays the beampattern contribution of RSMA-assisted DFRC without radar sequence in Scenario-UC with the same RMSE=$10$. We observe that the common stream of RSMA contributes more to target detection when the user pair with higher LB-IBR is served. This means that the common stream can further mitigate radar-to-communication interference compared to SDMA, and when such interference is larger, the common stream contributes even more and thus achieves a larger tradeoff gain (see for instance Fig. \ref{4b}  where the contribution of the common stream to the beampattern is significant, which leads to a larger tradeoff gain in Fig. \ref{5b}).
\subsubsection{RSMA Gain vs. Target's Location}
The second scenario denoted as Scenario-TL is set to serve the same two users, i.e., User-3 and User-4, and detect one target varying among all three prepared targets. \par 
\begin{figure}[t]
	\centering
	\subfigure[Tradeoffs when detecting Target-1.] {\includegraphics[width=0.48\linewidth]{t1_tradeoff.eps}\label{7a}}
	\subfigure[Tradeoffs when detecting Target-2.] {\includegraphics[width=0.48\linewidth]{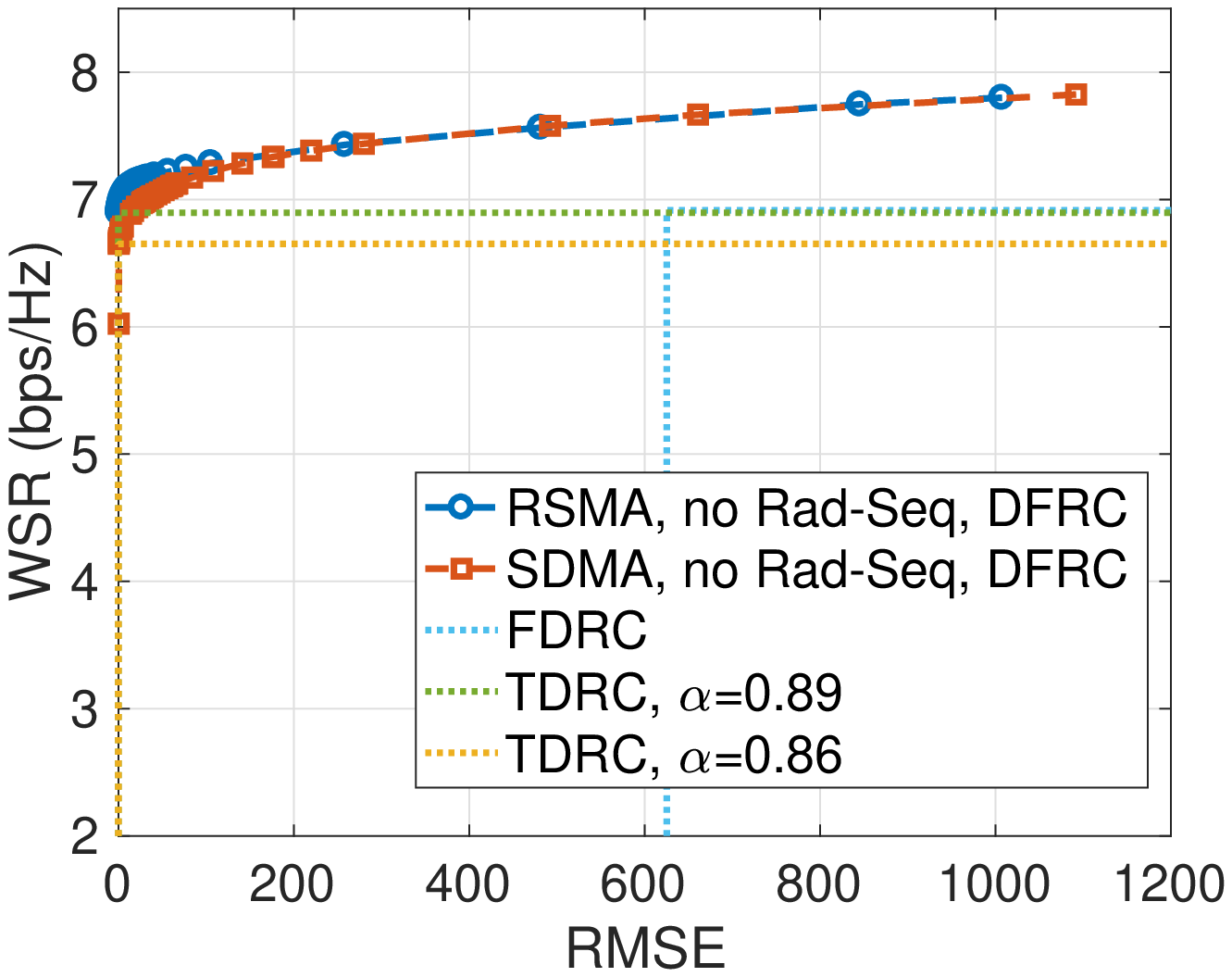}\label{7b}}
	\subfigure[Tradeoffs when detecting Target-3.] {\includegraphics[width=0.48\linewidth]{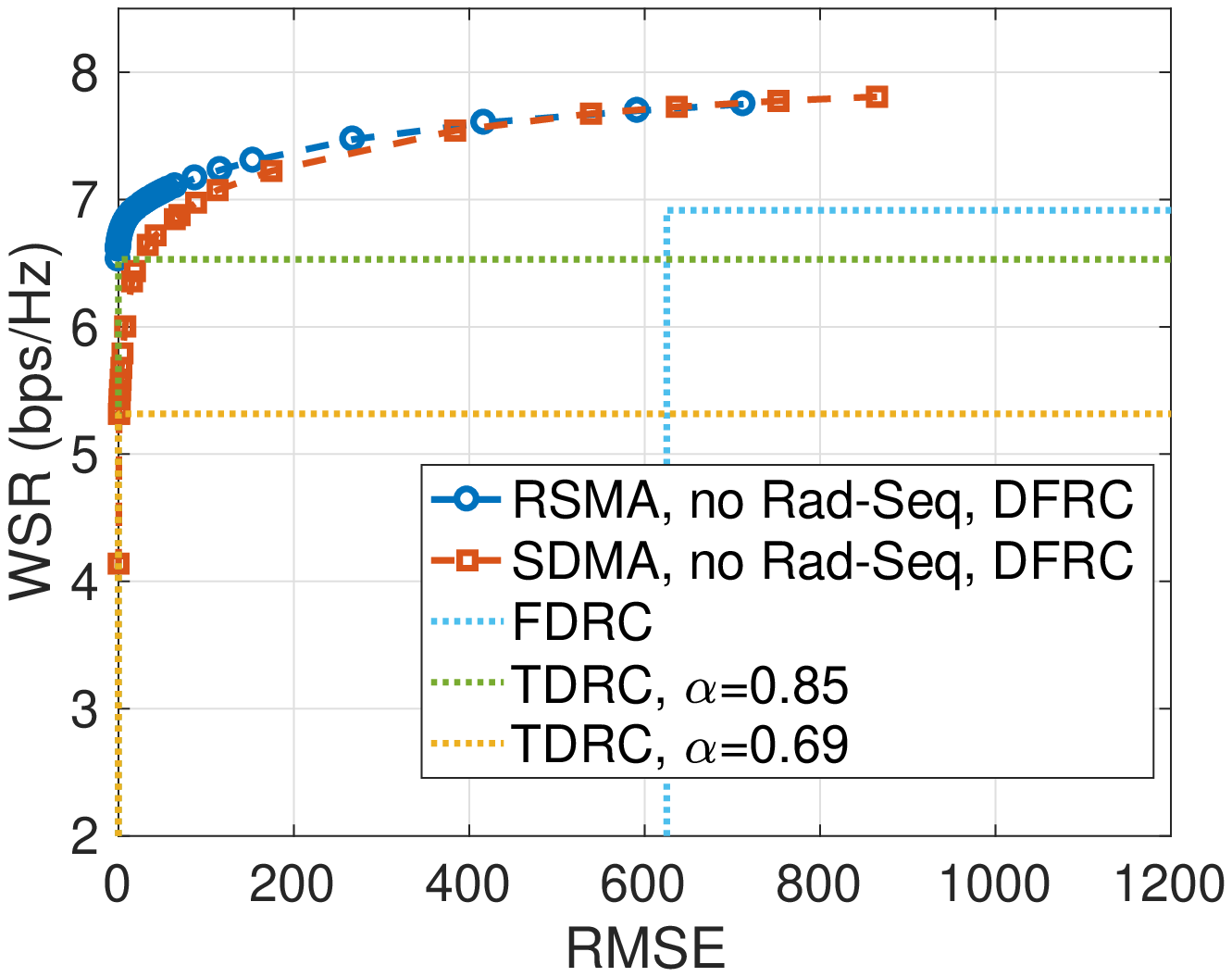}\label{7c}}
	\subfigure[LB-IBR for different target's locations.] {\includegraphics[width=0.48\linewidth]{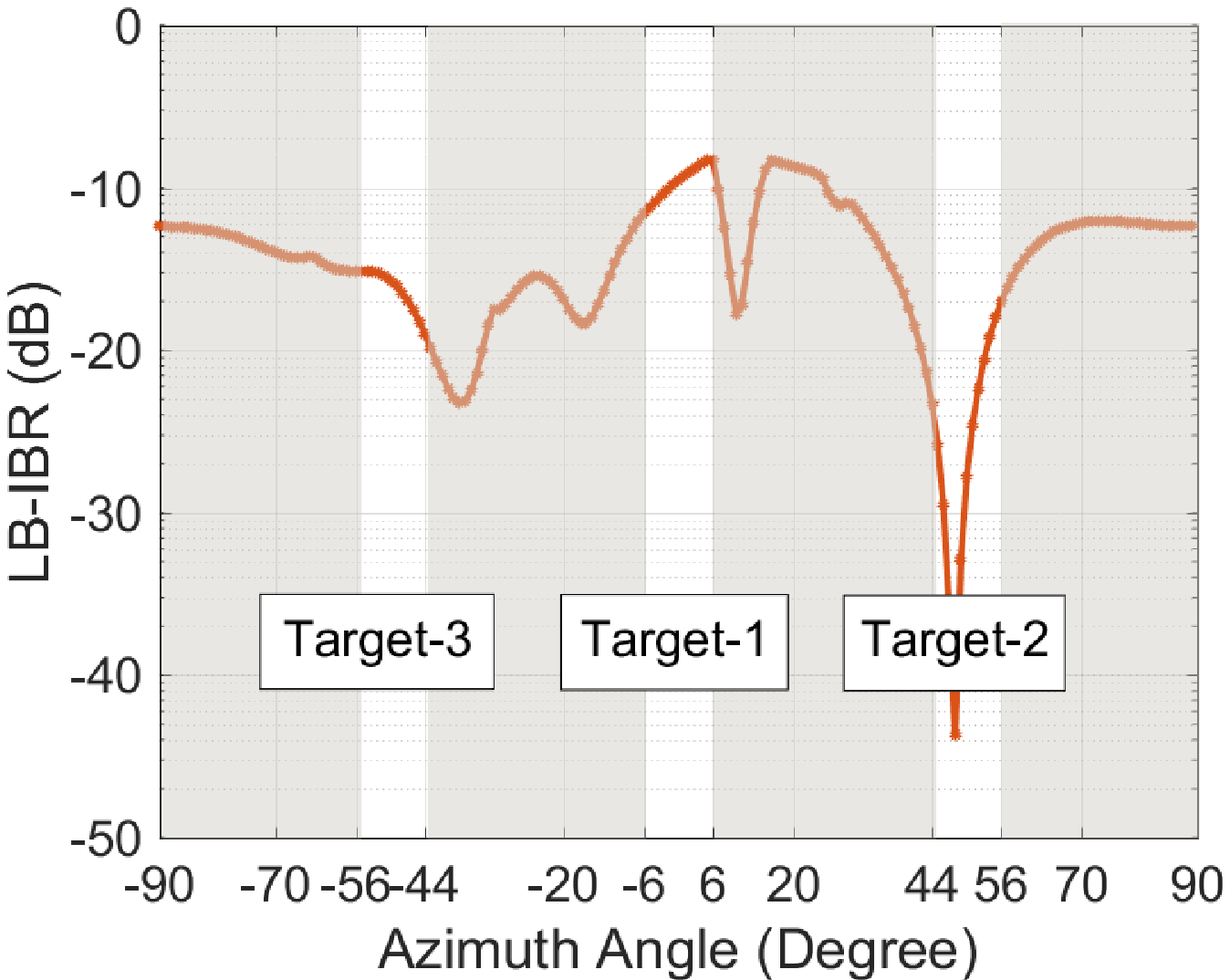}\label{TPIBR}}
	\caption{Tradeoff and LB-IBR for Scenario-TL.}\label{TP}
\end{figure}
Fig \ref{7a}--\ref{7c} first illustrate the tradeoffs in this scenario. Again, we see that RSMA-assisted DFRC without radar sequence outperforms FDRC as well as TDRC when $\alpha$ is smaller than around 0.8, and also has different tradeoff gains over SDMA-assisted DFRC without radar sequence. Likewise, we show in Fig. \ref{TPIBR} the LB-IBR at the different target locations. It shows again that RSMA helps to further mitigate the radar-to-communication interference.

\subsubsection{RSMA Gain vs. User Number}
The third scenario denoted as Scenario-UN aims at detecting Target-1, while serving a different number of users, i.e., two users (User-1,4), three users (User-1,2,4) and four users (User-1,2,3,4).\par
\begin{figure}[t]
	\centering
	\subfigure[Tradeoffs when serving two users.] {\includegraphics[width=0.48\linewidth]{2u_tradeoff.eps}}
	\subfigure[Tradeoffs when serving three users.] {\includegraphics[width=0.48\linewidth]{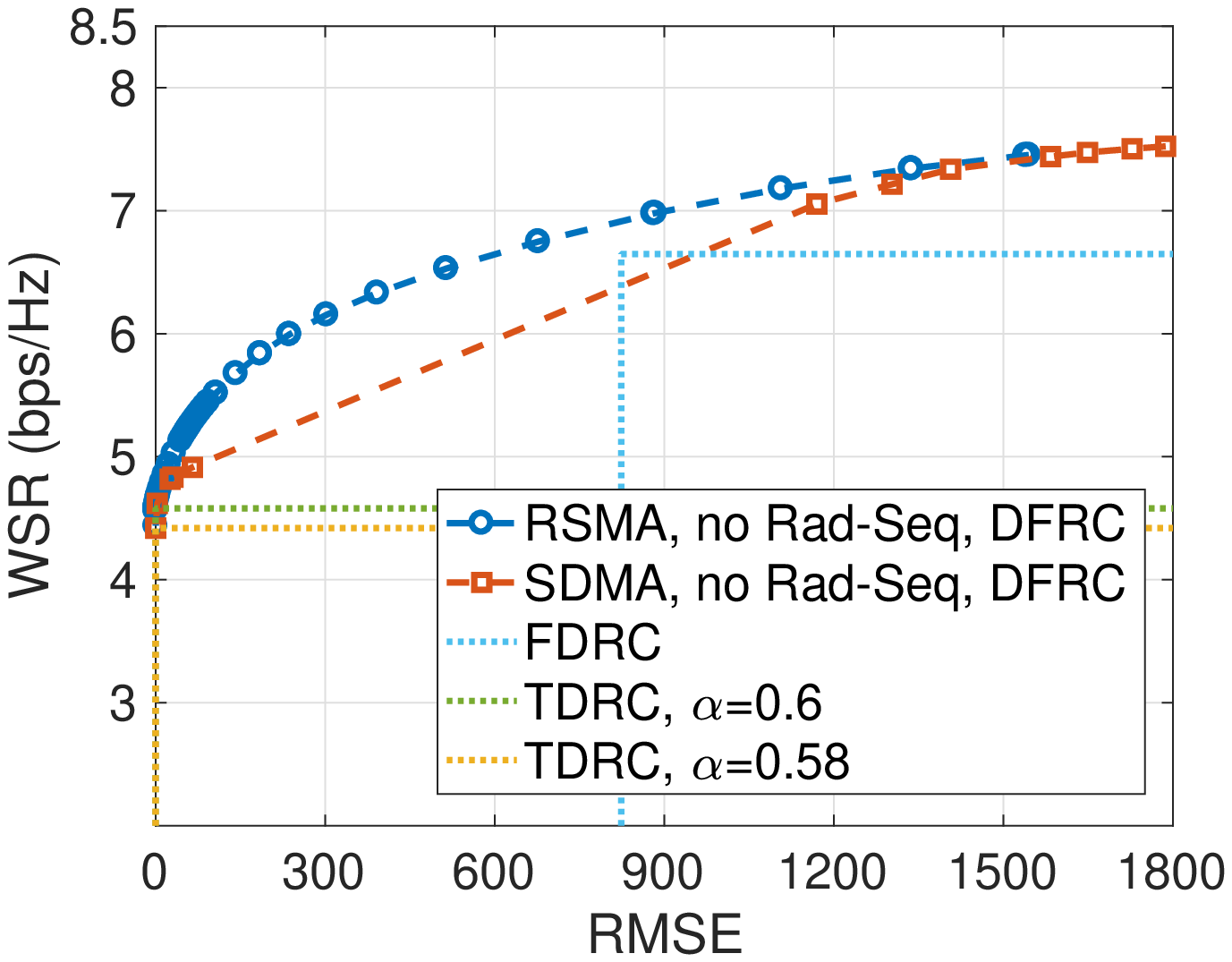}}
	\subfigure[Tradeoffs when serving four users.] {\includegraphics[width=0.48\linewidth]{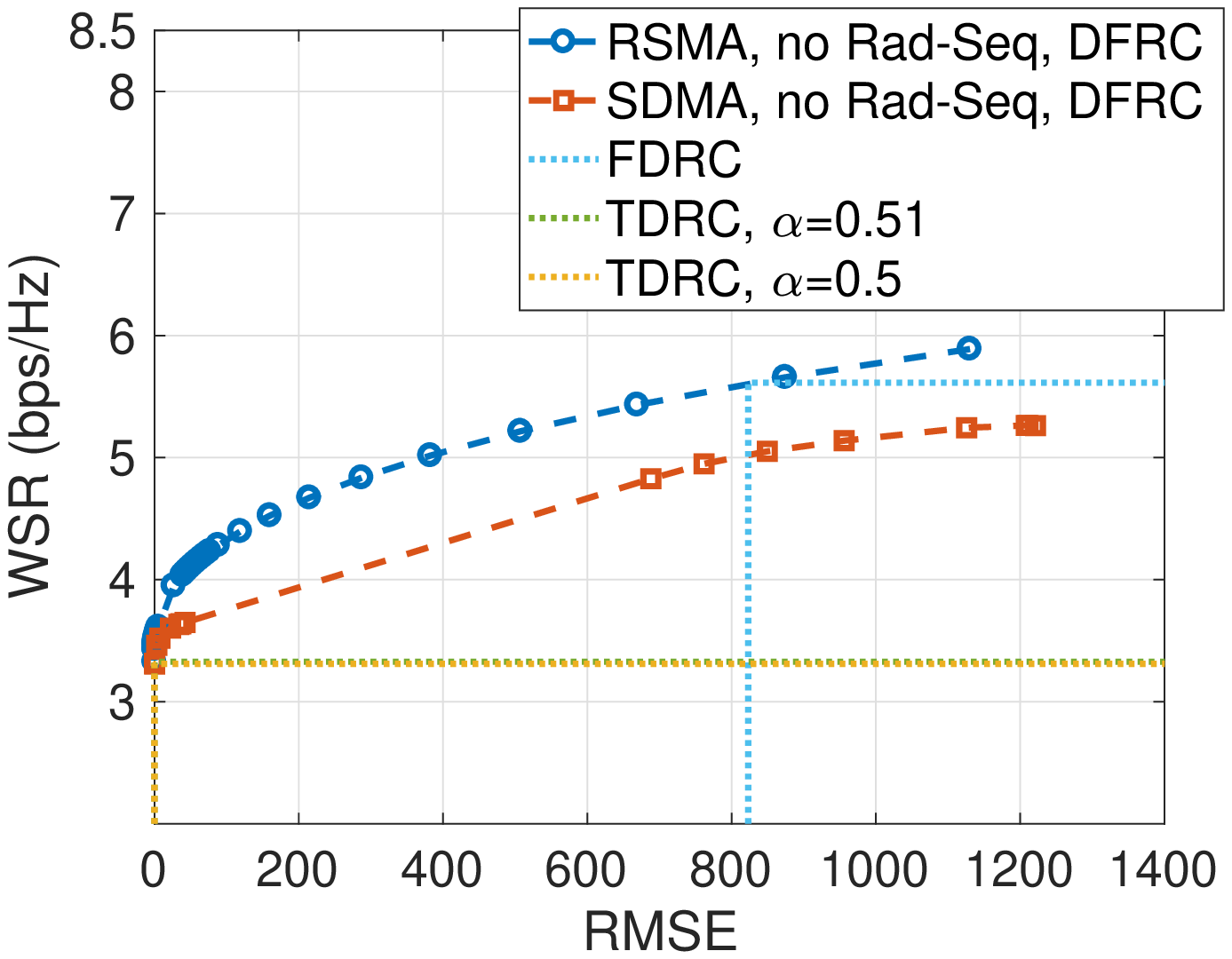}}
	\subfigure[LB-IBR for different user numbers.] {\includegraphics[width=0.48\linewidth]{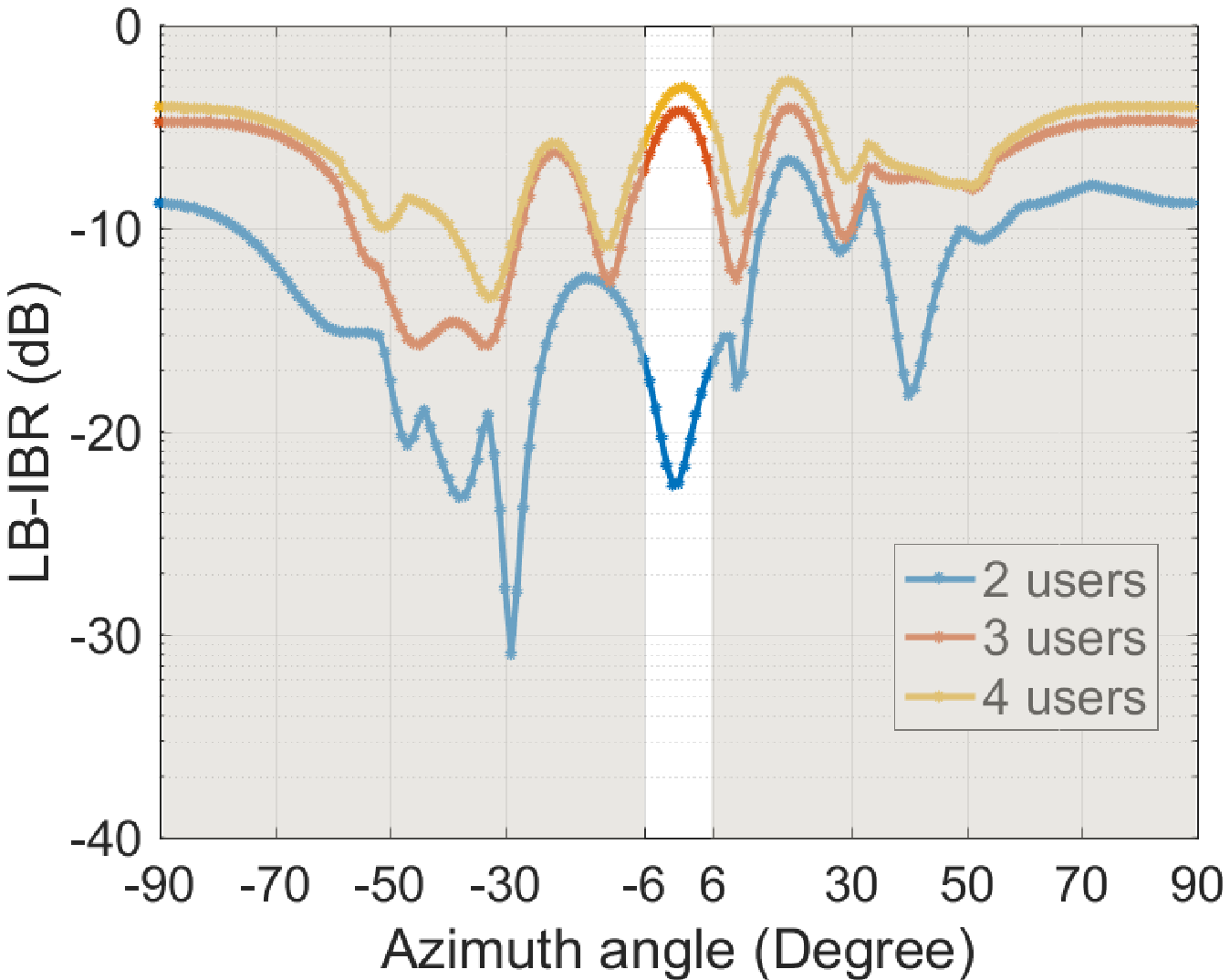}\label{UNIBR}}
	\caption{Tradeoff and LB-IBR comparison for Scenario-UN.}\label{UN}
\end{figure}

 
Fig. \ref{UN} shows the tradeoffs in all scenarios. Similarly, RSMA-assisted DFRC without radar sequence still outperforms FDRC, but when the number of user increases, it outperforms TDRC with smaller $\alpha$. It is also clear that RSMA-assisted DFRC without radar sequence has an advantage over SDMA-assisted DFRC without radar sequence. Fig. \ref{UNIBR} demonstrates the LB-IBR for this scenario, and again proves that RSMA brings more gain in the scenario with higher LB-IBR.\par  

\subsubsection{RSMA Performance in Multi-target Detection}
Next, we show the performance of RSMA to detect multiple targets. The scenario is denoted as Scenario-MUT where User-2 and User-4 are served and all three targets are detected.\par 
\begin{figure}[t]
	\begin{minipage}{0.48\linewidth}
		\centering
		\includegraphics[width=\linewidth]{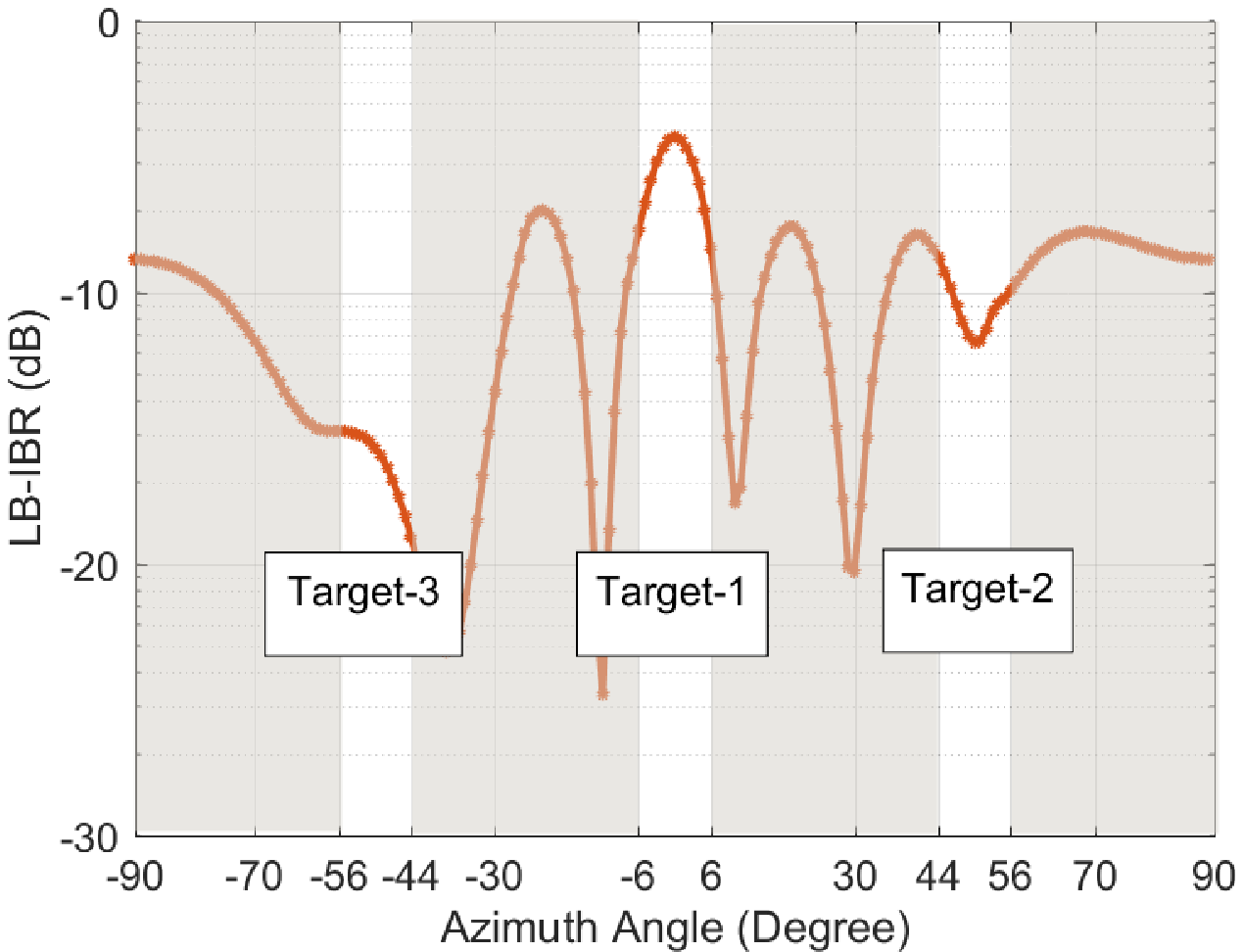}
		\caption{LB-IBR for Scenario-MUT.}\label{3t2u}
	\end{minipage}
	\begin{minipage}{0.48\linewidth}
	\centering
	\includegraphics[width=\linewidth]{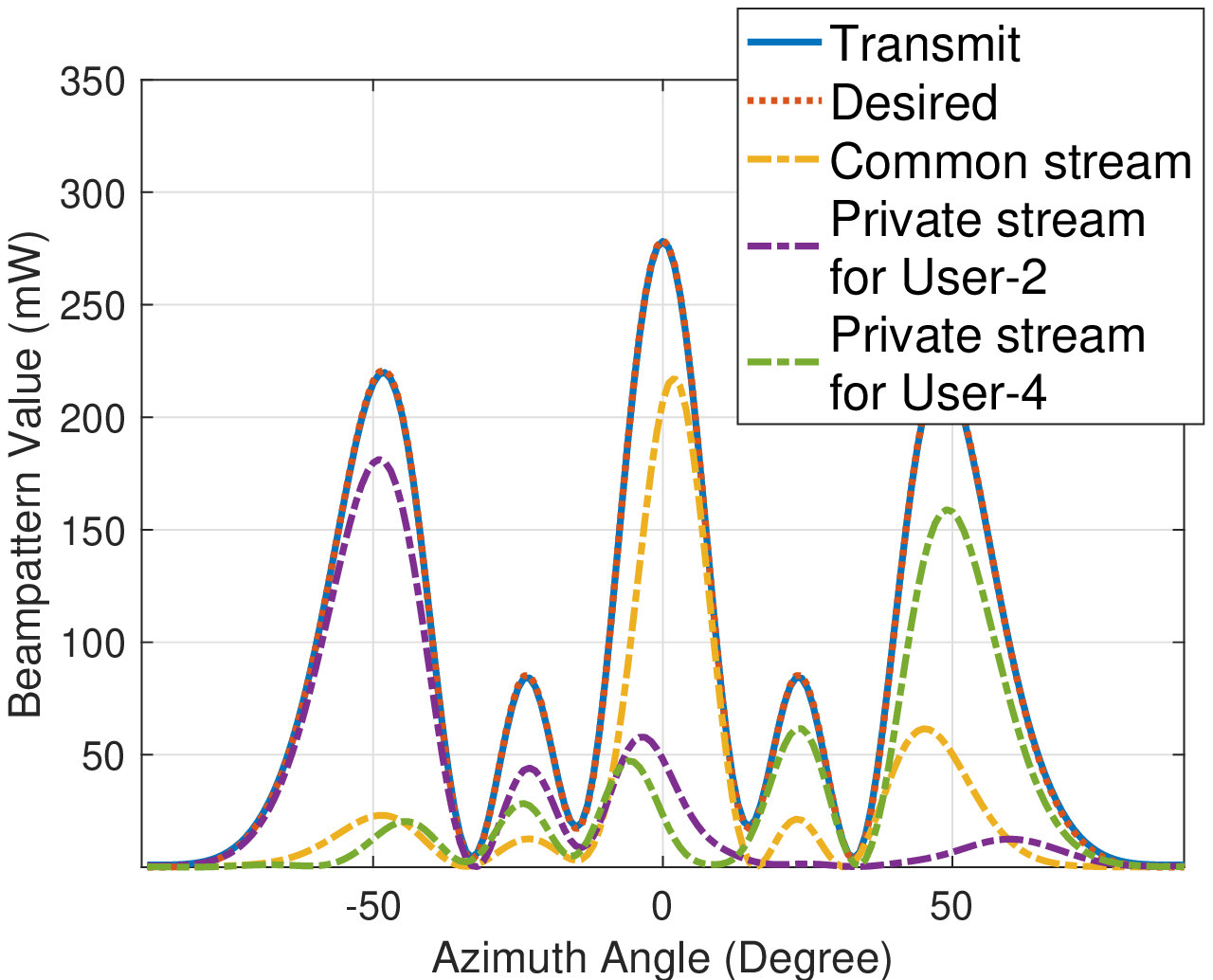}
	\caption{Beampattern contribution of RSMA-assisted DFRC without radar sequence for Scenario-MUT.}\label{3tbeamcon}
\end{minipage}
\end{figure}


Fig. \ref{3t2u} displays the LB-IBR in this scenario, where we can see LB-IBR is the highest at Target-1 but lowest at Target-3. We then show in Fig. \ref{3tbeamcon} the beampattern contribution of RSMA-assisted DFRC without radar sequence when $\text{RMSE}=10^1$, where the common stream contributes more to the target with higher LB-IBR.  \par 

All numerical experiments in this subsection show that RSMA-assisted DFRC without radar sequence outperforms FDRC, and in certain conditions surpasses TDRC. The common stream of RSMA also shows the capability of fulfilling the beampattern requirements of radar and mitigating radar-to-communication interference compared to SDMA in DFRC, which supports the observation that RSMA outperforms SDMA in DFRC.
\begin{figure}[t]
	\centering
\begin{minipage}{0.65\linewidth}
	\centering
	\includegraphics[width=\linewidth]{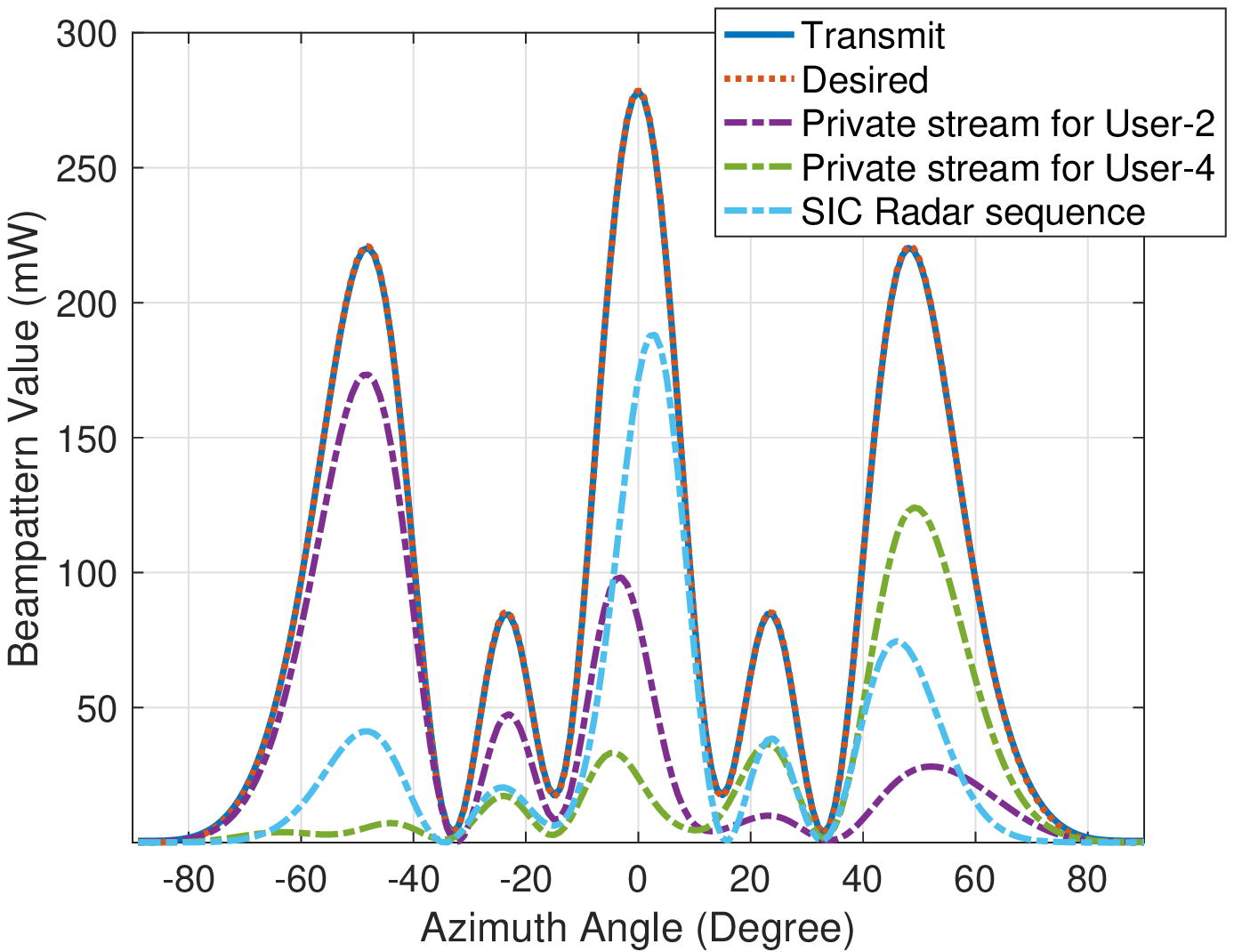}
	\caption{Beampattern contribution of SDMA-assisted DFRC with radar sequence and SIC for Scenario-MUT.}\label{3tbeamconMULPR1}
\end{minipage} 
	\begin{minipage}{0.65\linewidth}
	\centering
\includegraphics[width=\linewidth]{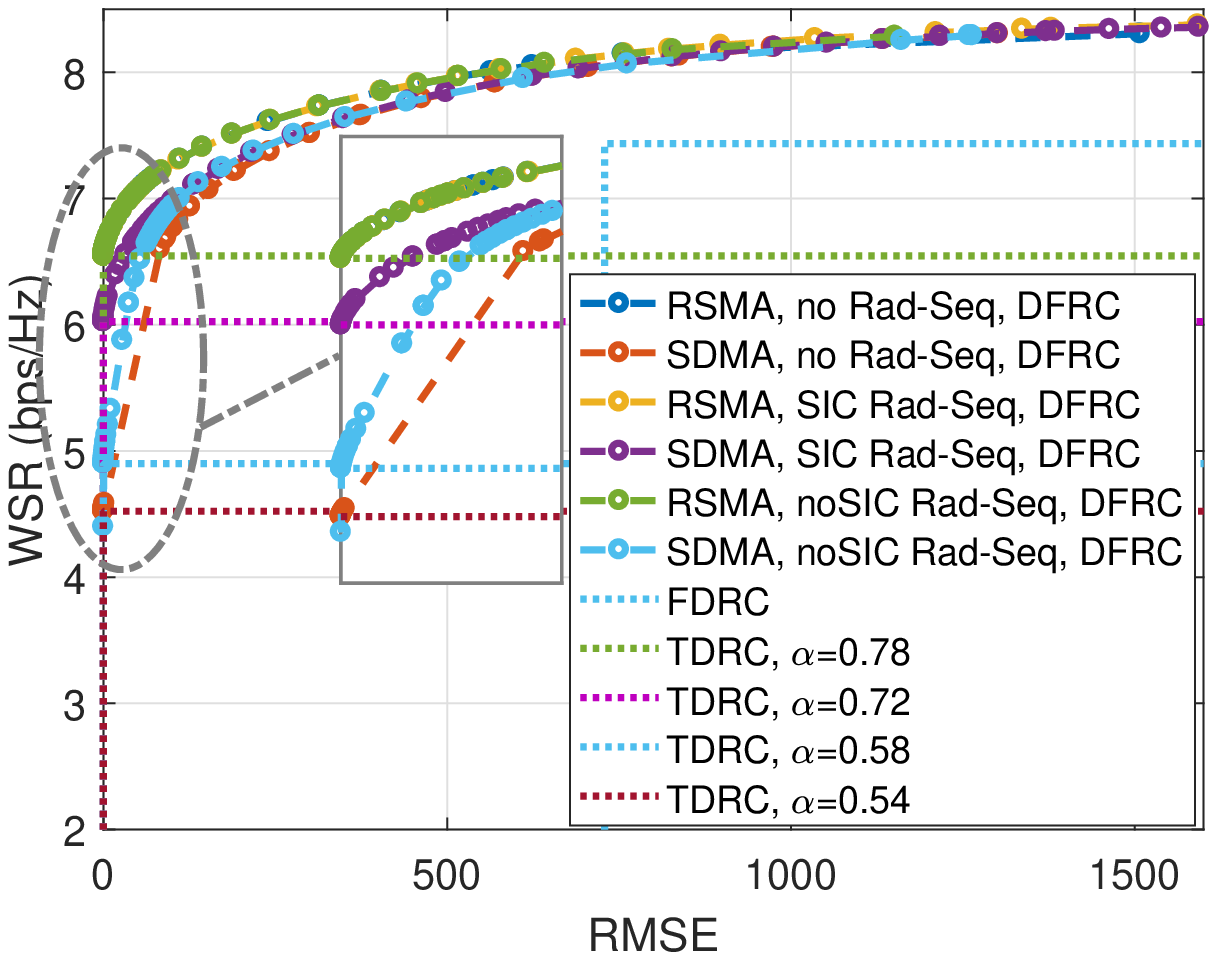}
\caption{Tradeoffs of all combinations of multiple access techniques and radar sequence modes for Scenario-MUT.}\label{full}
\end{minipage}
	\begin{minipage}{0.65\linewidth}
	\centering
\includegraphics[width=\linewidth]{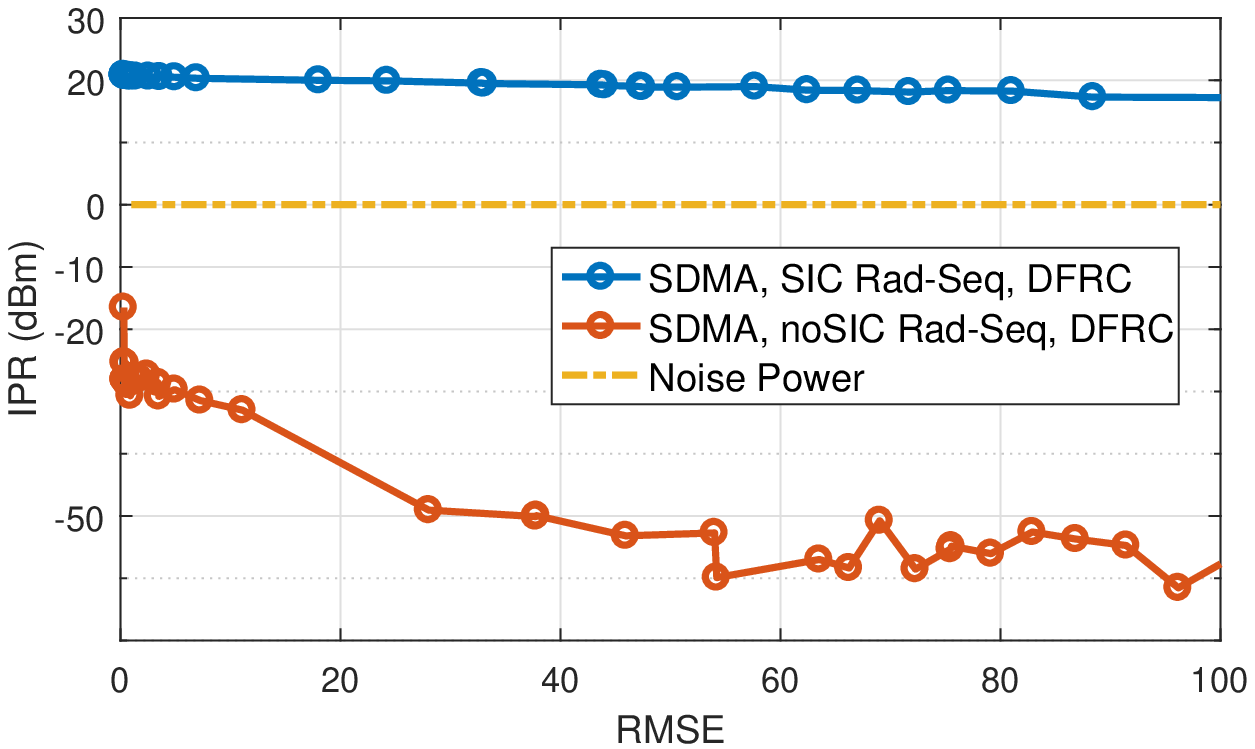}
\caption{IPR comparison between modes of SDMA-assisted DFRC with radar sequence.}\label{IPR}
\end{minipage}
\end{figure}
\subsection{RSMA vs. Radar Sequence in DFRC}
After investigating the advantage of enabling RSMA over SDMA in DFRC, we analyse DFRC's performance when radar sequence is enabled. \par 
Here, we adopt Scenario-MUT and obtain the tradeoffs of different combinations of multiple access techniques and radar sequence modes in Fig. \ref{full}.
It is clear that SDMA-assisted DFRC with radar sequence brings an around 0.4bps/Hz WSR gain over that without radar sequence when RMSE$\le1$, while enabling SIC to radar sequence leads to an additional 1.1bps/Hz WSR gain. However, when RSMA is used, it is interesting that whatever radar sequence mode is enabled, the tradeoff is the same with RSMA-assisted DFRC without radar sequence, which outperforms all SDMA-assisted DFRC systems, FDRC and TDRC when $\alpha<0.78$.\par 
We first explain in SDMA-assisted DFRC with radar sequence, why SIC brings benefits. We represent the total interference power of radar sequence imposed on communication users as $\text{IPR}=\sum_{k\in \mathcal{K}}\lVert{\bf h}^H_k{\bf p}_{\text{r}}\lVert_2$. In Scenario-MUT, IPRs of both modes of SDMA-assisted DFRC with radar sequence are demonstrated corresponding to RMSE in Fig. \ref{IPR}. We can see that when SIC is disabled, IPR is significantly low after optimization, which indicates that ${\bf p}_{\text{r}}$ is designed obeying ZF to mitigate interference as much as possible. In contrast, when SIC is used, IPR is comparatively high, because it can always be canceled via SIC and thus ${\bf p}_{\text{r}}$ can be designed freely without obeying ZF. Since the restriction on ${\bf p}_{\text{r}}$ is much weaker in SDMA-assisted DFRC with radar sequence and SIC, the feasible region is enlarged. \par 
We then illustrate why RSMA outperforms SDMA-assisted DFRC with radar sequence. Fig. \ref{3tbeamconMULPR1} illustrates the beampattern contribution of SDMA-assisted DFRC with radar sequence and SIC in Scenario-MUT with RMSE=$10^1$, where radar sequence acts quite similarly compared to the common stream in RSMA-assisted DFRC without radar sequence in Fig. \ref{3tbeamcon}. Despite the same beampattern behaviour, RSMA-assisted DFRC without radar sequence still has a higher WSR. This is because SIC to the radar sequence is not used for managing the multiuser communications, but only the interference between radar and communication. However, apart from fulfilling the radar beampattern and managing the radar-to-communication interference, the common stream in RSMA can further manage the interference among communication users as it can be decoded partly at users. Therefore, the SIC receiver in RSMA is used for the dual purpose of managing interference between communication users but also the interference between communication and radar.\par 
To summarize, the common stream of RSMA has the triple function of 1) managing interference between communication users, 2) managing interference between communication and radar, and 3) fulfilling the beampattern requirements of the radar. As a consequence, there is no need for an additional radar sequence and SIC. This coincides with the observation that RSMA-assisted DFRC with and without radar sequence achieve the same tradeoff performance.

\subsection{Convergence}
\begin{figure}[t]
   	\centering
   	\includegraphics[width=0.65\linewidth]{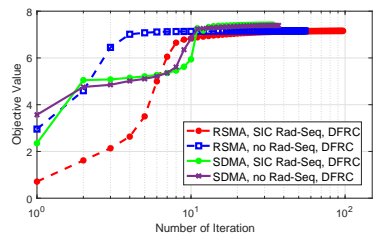}
   	\caption{Convergence performance of the ADMM framework for solving the DFRC design problem, $\lambda=10^{-5}$.}\label{converge}
   \end{figure}

Next, we show the convergence performance of the proposed ADMM framework solving the DFRC design problem. We consider the scenario where User-2 and User-4 are served while Target-1 is to be detected. In this scenario, Fig. \ref{converge} shows the convergence of the ADMM framework that solves the optimization problem of designing SDMA-assisted DFRC without radar sequence, RSMA-assisted DFRC without radar sequence, SDMA-assisted DFRC with radar sequence and SIC, and RSMA-assisted DFRC with radar sequence and SIC. We set $\lambda=10^{-5}$ in \eqref{CR origin} in Fig. \ref{converge}. It is clear that the convergence of the objective values of the proposed ADMM framework is guaranteed, and the framework converges to a good solution in around ten times of iteration. 
\section{Conclusion}
To conclude, we propose a multi-antenna DFRC system that enables RSMA and different modes of radar sequence. The message split, precoders of information streams and radar sequence are optimized together with the aim of jointly maximizing WSR and minimizing MSE of beampattern approximation under the average transmit power constraint at each antenna. We propose the ADMM-based framework to effectively solve the intractable non-convex optimization problem of the proposed RSMA-assisted DFRC. SDMA-assisted DFRC, practically simpler TDRC and FDRC are compared as baselines. The DFRC tradeoff between WSR and beampattern approximation MSE is investigated. Numerical results show that without radar sequence, RSMA-assisted DFRC has a better tradeoff performance than SDMA-assisted DFRC, DFRC and TDRC in certain conditions. Most importantly, although enabling radar sequence improves the performance of SDMA-assisted DFRC and its interference on communication can be removed by SIC,  RSMA-assisted DFRC without radar sequence even outperforms SDMA-assisted DFRC with radar sequence and SIC, and shares the same performance with RSMA-assisted DFRC with radar sequence and SIC. This is because the common stream of RSMA not only has the same function with radar sequence to match the transmit beampattern and manage the interference between radar and communication, but also has the unique capability to manage the interference among communication users. The SIC receiver of RSMA is thus better exploited to manage the interference between radar and communication as well as among communication users. As a result, by using RSMA, the DFRC performance is enhanced but the architecture is simplified because there is no need to use an additional radar sequence and SIC. We conclude that RSMA is a more powerful and promising technique for DFRC.

\appendix[Proof of \eqref{LBIBR}]
We denote ${\bf p}'=\left[{\bf p}_1^T, {\bf p}_2^T, \dots, {\bf p}_K^T\right]^T$, and the selection matrix in \eqref{DPK} as
$
{\bf D}_{\text{p},k}=\left[{\bf 0}^{N_{\text{t}}\times(k-1)N_{\text{t}}}\quad{\bf I}_{N_{\text{t}}}\quad {\bf 0}^{N_{\text{t}}\times(K-k)N_{\text{t}}}\right],
k=1,2,\dots,K.
$
Then, IBR$(\theta_m)$ is rewritten as
\begin{equation}
\begin{split}
\text{IBR}(\theta_m)
&=\frac{{\bf p}'^H\left(\sum_{k\in \mathcal{K}}\sum_{j\neq k}{\bf D}_{\text{p},k}^H{\bf h}_j{\bf h}_j^H{\bf D}_{\text{p},k}\right){\bf p}'}{{\bf p}'^H\left(\sum_{k\in \mathcal{K}}{\bf D}_{\text{p},k}^H{\bf a}(\theta_m){\bf a}(\theta_m)^H{\bf D}_{\text{p},k}\right){\bf p}'}.
\end{split}
\end{equation}
Denoting ${\bf M}=\sum_{k\in \mathcal{K}}{\bf D}_{\text{p},k}^H{\bf a}(\theta_m){\bf a}(\theta_m)^H{\bf D}_{\text{p},k}$ and ${\bf L}=\sum_{k\in \mathcal{K}}\sum_{j\neq k}{\bf D}_{\text{p},k}^H{\bf h}_j{\bf h}_j^H{\bf D}_{\text{p},k}$, we have 
$
\text{IBR}(\theta_m)=\frac{{\bf p}'^H{\bf L}{\bf p}'}{{\bf p}'^H{\bf M}{\bf p}'}
$.\par 
According to the property of Rayleigh-Ritz quotient, we have the following inequality
$	\frac{{\bf p}'^H{\bf L}{\bf p}'}{{\bf p}'^H{\bf M}{\bf p}'}\geq \lambda_{\text{min}}\left({\bf M}^{-1}{\bf L}\right)$
where $\lambda_{\text{min}}$ is the minimum eigenvalue of the matrix respectively. Since ${\bf M}$ does not have full rank, we substitute the inverse with the pseudo inverse ${\bf M}^{\dag}$. With the fact that 
$
\frac{1}{N_{\text{t}}}{\bf M}{\bf M}={\bf M},
$
we have ${\bf M}^{\dag}=(1/N_{\text{t}}^2){\bf M}$. Then we get ${\bf M}^{\dag}{\bf L}=(1/N_{\text{t}}^2){\bf M}{\bf L}=(1/N_{\text{t}}^3){\bf M}{\bf M}{\bf L}$. Since the eigenvalues of ${\bf M}{\bf M}{\bf L}$ are the same with those of ${\bf M}{\bf L}{\bf M}$, we further derive
\begin{equation}
\begin{split}
{\bf M}&{\bf L}{\bf M}\\
=&\frac{1}{N_{\text{t}}^3}\sum_{i\in \mathcal{K}}\sum_{k\in \mathcal{K}}\sum_{j\neq k}\sum_{i'\in \mathcal{K}}\left[{\bf D}_{p,i}^H{\bf a}(\theta_m){\bf a}(\theta_m)^H{\bf D}_{p,i}{\bf D}_{\text{p},k}^H{\bf h}_j{\bf h}_j^H\right.\\
&\left.{\bf D}_{\text{p},k}{\bf D}_{p,i'}^H{\bf a}(\theta_m){\bf a}(\theta_m)^H{\bf D}_{p,i'}\right]\\
=&\frac{1}{N_{\text{t}}^3}\sum_{k\in \mathcal{K}}\left[\sum_{j\neq k}\lVert{\bf a}(\theta_m)^H{\bf h}_j\lVert_2^2\right]{\bf D}_{\text{p},k}^H{\bf a}(\theta_m){\bf a}(\theta_m)^H{\bf D}_{\text{p},k}.
\end{split}
\end{equation} 
It is easy to get the $K$ eigenvalues of ${\bf M}^{\dag}{\bf L}$ as $\lambda_k=(1/N_{\text{t}}^2)\sum_{j\neq k}\lVert{\bf a}(\theta_m)^H{\bf h}_j\lVert_2^2$. Then, we have
$\lambda_{\text{min}}\left({\bf M}^{\dag}{\bf L}\right)=(1/N_{\text{t}}^2)\min_{k\in \mathcal{K}}(\sum_{j\neq k}\lVert{\bf a}(\theta_m)^H{\bf h}_j\lVert_2^2)$.
Thus, \eqref{LBIBR} is proved.

%

%
%
%
%
%

\ifCLASSOPTIONcaptionsoff
  \newpage
\fi



\bibliographystyle{IEEEtran}
\bibliography{xcc}
\end{document}